\documentclass[MSNbibl,nameyear,seceqn,dvips]{arxstspdf}
\usepackage{graphicx}
\usepackage{flushend}
\usepackage{stfloats}

\volume{26}
\issue{1}
\pubyear{2011}
\firstpage{116}
\lastpage{129}
\doi{10.1214/11-STS353}

\makeatletter

\makeatletter

\def\conD{\stackrel{d} {\longrightarrow}}
\def\var{\operatorname{Var}}
\def\diag{\operatorname{diag}}
\def\cov{\operatorname{Cov}}
\def\logit{\operatorname{logit}}
\def\bmu{{\boldsymbol{\mu}}}
\def\bSigma{{\boldsymbol{\Sigma}}}

\makeatother

\begin{document}
\begin{frontmatter}

\title{Statistical Analysis in Genetic Studies of Mental Illnesses}

\runtitle{Genetics of Mental Illnesses}

\begin{aug}
\author{\fnms{Heping} \snm{Zhang}\corref{}\ead[label=e1]{heping.zhang@yale.edu}}

\runauthor{H. Zhang}

\affiliation{Yale School of Public Health}

\address{Heping Zhang is Professor, Yale School of Public Health, Yale University, 60
College Street, New Heaven, Connecticut 06520-8034, USA \printead{e1}.}

\end{aug}

% ABSTRACT
%
\begin{abstract}
Identifying the risk factors for mental illnesses is of significant
public health importance. Diagnosis, stigma associated with mental
illnesses, comorbidity, and complex etiologies, among others, make it
very challenging to study mental disorders. Genetic studies of mental
illnesses date back at least a century ago, beginning with descriptive
studies based on Mendelian laws of inheritance. A variety of study
designs including twin studies, family studies, linkage analysis, and
more recently, genomewide association studies have been employed to
study the genetics of mental illnesses, or complex diseases in general.
In this paper, I will present the challenges and methods from a
statistical perspective and focus on genetic association studies.
\end{abstract}

% KEYWORDS
%

\end{frontmatter}

%s1 ###
\section{Introduction} \label{2010-02-24-1}

Mental illnesses affect the health and well-being of all populations
and all ages. Schizophrenia---a~chro\-nic, severe, and disabling brain
disorder---is one of these mental illnesses, affecting about 1.1
percent of the U.S. population age 18 and older in a given year. People
with schizophrenia sometimes hear voices others do not hear, believe
that others are broadcasting their thoughts to the world, or become
convinced that others are plotting to harm them. These experiences can
make them fearful and withdrawn and cause difficulties when these
people try to have relationships with others
(\href{http://www.nimh.nih.gov}{http://www.nimh.nih.gov}). Emil Kraepelin (1856--1926) described
``Dementia Praecox'' as an inherited disorder in his influential
``Textbook of Psychiatry'' (\citeyear{Kra99}). Dementia Praecox, coined
``schizophrenia,'' was first used by Arnold Pick (1851--1924)---a
professor of psychiatry at the German branch of Charles University in
Prague---to describe a patient with a psychotic disorder resembling
hebephrenia in 1891.

Nearly a century ago, Cannon and Rosanoff (\citeyear{CanRos11}) made an attempt to
understand whether there are any forms of nervous and mental diseases
that are transmitted from generation to generation in concordance with
Mendelian laws. They examined the families of 11 neuropathetic
patients, which are now referred to as probands in pedigrees. Using
Mende\-lian laws as their theoretical expectation, they concluded that
the neuropathetic make-up is recessive to normal. Although the report
was indeed ``preliminary,'' a few things are noteworthy. First, they
noted that ``any form of insanity or even all the forms of hereditary
insanity do not constitute an independent hereditary character.'' This
raised an early sign of the complexity associated with studying mental
disorders compared to the characterization of the disorders and their
comorbidity. Here, comorbidity refers to more than one disease
condition in the same patient. Second, they remarked ``should larger
accumulations of such data in the future give similar results, we shall
be able'' to confirm their result. The requirement for more samples and
replication is another challenge in studies of complex diseases. Last,
but not the least, while they said ``let us test$,\ldots,$ the
hypothesis $\ldots$'' they did not mean a statistical test. However,
the idea of the $\chi^2$-test is evident.

Despite this early work, it was not until the 1960s that the
researchers began to use scientifically rigorous designs and methods to
study the inheritance of mental illnesses. For example, the key idea in
adoption studies lies in the belief that any links between an adopted
child and the biological parents are attributable to genetics, and any
links between that child and adoptive parents can be attributed to
environment (Plomin et al., \citeyear{Ploetal97}). This enables us to separate the
confounding environment (i.e., a family) from genetic contribution.
Consequently, there are two strategies in adoption studies. One
approach compares the risk of developing schizophrenia in the adopted
children of schizophrenic parents to the risk of adopted children whose
parents do not have schizophrenia. Several studies including Heston
(\citeyear{Hes66}), Rosenthal (\citeyear{Ros72}) and Tienari (\citeyear{Tie91}) used this approach to study
schizophrenia. Each study found an elevated risk in adopted-away
children of schizophrenic parents, supporting the role of genetics in
the transmission of schizophrenia. The origin of this approach is the
schizophrenic parents. Another approach backtracks from adopted
children who have developed schizophrenia and compares the risks of
schizophrenia in their adoptive and biological families. Kety, Rosenthal and
Wender
(\citeyear{KetRosWen78}) and others found that the risk was significantly higher in the
biological relatives than in the adoptive families, again underscoring
the role of genetics as a risk factor.

While these schizophrenia adoption studies are influential in
understanding the role of genetics in mental disorders, the majority of
the genetic factors associated with mental disorders are based on
family and twin studies. By comparing the concordance in the risk
between identical (monozygotic) and fraternal (dizygotic) twins, twin
studies arguably provide the most compelling results about genetic and
environmental effects. For example, the concordance in monozygotic
twins for Tourette's syndrome, a~complex disorder characterized by
repetitive, sudden and involuntary movements or noises called tics, was
reported to be about 50\% whereas it is less than 10\% in dizygotic
twins.

Twin studies are most helpful in demonstrating the magnitude of genetic
effect, but they do not provide insight into the inheritance pattern of
a condition. Thus, family studies can offer information that twin
studies cannot. Thus, Cannon and Rosanoff (\citeyear{CanRos11}) employed a
small-scale, simple family study. Using the Mendelian laws, not only
might we find evidence of genetics, but also infer the mode of
transmission, as Cannon and Rosanoff (\citeyear{CanRos11}) concluded for the heredity
of insanity.

Although twin and family studies continue to be useful for
understanding the genetics of complex diseases, different studies are
needed to locate a specific gene on a chromosome that may underlie the
disease. Gene mapping in humans through linkage analysis emerged in the
1930s, but it was Morton (\citeyear{MOR55}) who laid the foundation for the
methodology. It was only during the 1970s and 1980s, when the
Elston--Steward (\citeyear{ElsSte71}) algorithm was developed and implemented (Ott,
\citeyear{Ott74}), that the method thrived as a~common tool of genetic studies.
These initial and subsequent developments allowed for linkage analyses
of multiple markers simultaneously. In light of the sheer number of
genes and that we do not know which specific gene we are looking for,
we typically genotype 300 to 400 ``landmarks'' that cover the 22 pairs
of autosomes and the X chromosome. By inferring the transmission
patterns of these markers, then linking them to the disease status, we
can obtain information about the most probable region where the gene of
interest resides.

$\!\!$While linkage studies have had some successes (e.g., BRCA1), they have
generated many more premature excitements. In the late 1980s, two
particular studies attracted significant public attention after they
reported that bipolar affective disorders were linked to DNA markers on
chromosome 11, and that a susceptibility locus for schizophrenia was
located on chromosome 5. Unfortunately, these findings were not
replicated. Replications in genetic studies of mental disorders do not
come easily. For example, Abelson et al. (\citeyear{Abeetal05}) identified mutations
involving the SLITRK1 gene (13q31.1) in a small number of people with
Tourette's syndrome. However, most people with Tourette's syndrome do
not have a mutation in the SLITRK1 gene. Because the mutations were
reported in so few people with this condition, the association of the
SLITRK1 gene with this disorder could not be confirmed. In fact, Scharf
et al. (\citeyear{Schetal08}) reported a lack of the association between SLITRK1var321
and Tourette's syndrome in a large family-based sample.

Various reasons have been suggested to explain the difficulties
detouring progress in genetic studies using linkage analysis. A key
concept underlying linkage analysis is the recombination fraction,
which reflects the distance between any two markers, such as a DNA
marker and the disease locus. There may be limited information in the
data, however, diminishing the power of the linkage study.
Furthermore, complex diseases are polygenic, involving multiple genes
(Carter and Chung, \citeyear{CarChu80}). Linkage analyses, however, are generally
under the assumption of one major gene. Additionally, heterogeneity in
the diagnosis and comorbidity of mental illnesses make linkage analysis
considerably more difficult, if even possible at all.

Many investigators have adopted association analyses to take advantage
of the advent of high-through\-put genotyping technologies. Recent
efforts have\break identified genes that contribute to a number of complex
human traits using the ultra-dense genetic mar\-kers (Arking et al.,
\citeyear{Arketal06}; Klein et al., \citeyear{Kleetal05}; Duerr et al., \citeyear{Dueetal06}; Chen et al., \citeyear{Cheetal07}). Trios
(one affected offspring and two parents) have been an effective design
for association studies, particularly with the development of the
elegant transmission/disequili\-brium test (TDT) (Spielman, McGinnis and
Ewens, \citeyear{SpiMcGEwe93}). The central idea of this test is that each affected child
serves as his or her own matched case and control. This acts to control
for all potential confounding issues and examines alleles that both are
and are not transmitted from the parents. In the absence of association
between the affective status and the gene, the distributions of the
transmitted and non-transmitted alleles are expected to be the same.
Deviations in distribution as evaluated by a $\chi^2$-test indicate the
existence of association. Trios are the simplest example of nuclear
family, but when other siblings are available, the trio design is not
cost-effective. As a result, family-based association tests (FBAT)
including sibships (Spielman and Ewens, \citeyear{SpiEwe98}; Horvath and Laird, \citeyear{HorLai98};
Knapp, \citeyear{Kna99}), nuclear families (Weinberg, \citeyear{Wei99}; Lunetta et al., \citeyear{Lunetal00};
Rabinowitz and Laird, \citeyear{RabLai00}) and general pedigrees (Martin, Monks,
Warren and Kaplan, \citeyear{Maretal00}) have been developed.

Another restriction in the use of trios is the requirement of defining
the affective status of a disease. Consequently, association tests have
been proposed for quantitative traits (Allison, \citeyear{All97}; Rabinowitz,
\citeyear{Rab}), traits with distribution belonging to an exponential family
(Liu, Tritchler and Bull, \citeyear{LiuTriBul02}), ordinal traits (Zhang, Wang and Ye,
\citeyear{ZhaWanYe06}; Wang, Ye and Zhang, \citeyear{WanYeZha06}) and multiple traits (Lange et al.,
\citeyear{Lanetal03}; Zhang, Liu and Wang, \citeyear{ZhaLiuWan10}).

Since the early success in identifying the complement factor H
polymorphism in age-related macular degeneration (Klein et al., \citeyear{Kleetal05}),
case-control association studies have intensified, and many genetic
variants have been identified and catalogued (Hindorff et al., \citeyear{Hinetal09}).
Despite the enormous investment, the intense attention to the genetics
of diseases, the rapid improvement in technology, and the increasingly
large sample sizes in many studies, it remains challenging to identify
disease genes, especially those underlying mental illnesses. Some of
the common genetic variants that have been identified for complex
diseases only account for a small portion of the genetic risk, which
may vary across populations (Goldstein, \citeyear{Gol09}). For example, Kopp et al.
(\citeyear{Kopetal08}) and Kao et al. (\citeyear{autokey33}) identified several variations in the MYH9
gene as major contributors to excess risk of kidney disease among
African-Americans. They found that 60 percent of African-Americans
carry the risk variants as opposed to 4 percent of white Americans.

$\!\!\!$Technology will continue to improve and the amount of genetic data will
increase. The purpose of this article is to review some of the progress
from a statistical perspective and discuss some of the potential
challenges. Obviously, it would take volumes or series to do justice to
all of the work in statistical genetics. Instead of taking on that
impossible task, this article is oriented toward the publications
directly related to my own recent work.

%s2 ###
\section{Methods} \label{2010-02-24-2}

Since 1952, the American Psychiatric Association has published four
editions of the Diagnostic and Statistical Manual of Mental Disorders
(DSM) and plans to release its fifth edition in 2013. While widely
used, the use and development of the DSM has not gone without
controversy and criticism. Unlike diseases for which the diagnoses are
well accepted by physicians and patients, such as cancer, the diagnosis
of mental disorders must reflect biological factors (e.g., gender and
racial disparities), non-biological factors such as culture that are
not specific to one person, and it also must reflect the natural
variation within the same person.

%s2.1 ###
\subsection{Ordinal Traits} \label{2010-02-24-3}

It is clear from the above discussion that a simple dichotomous
diagnosis (e.g., yes or no), or a well-distributed continuous trait, is
unlikely to characterize the state of mental disorders. In fact, the
questions used in the diagnosis of mental disorders, such as DSM-IV,
are usually posed in terms of severity or frequency, and hence in an
ordinal scale.

Statistical methods for genetic analysis are well established for both
quantitative (continuous) and binary traits (see, e.g., Blackwelder and
Elston, \citeyear{BlaEls85}; Goldgar, \citeyear{Gol90}; Schork, \citeyear{Sch93}; Amos, \citeyear{Amo94}; Risch and Zhang,
\citeyear{RisZha95}; Kruglyak et al., \citeyear{Kruetal96}; Blangero and Almasy, \citeyear{BlaAlm97}; Ott, \citeyear{Ott99}).
While there has been some progress in the analysis of ordinal traits
(e.g., Heath et al., \citeyear{Heaetal02}; Steinke, Borish and
Rosenwasser, \citeyear{SteBorRos03}; Vergne et al., \citeyear{Veretal03};
Zhang, Feng and Zhu, \citeyear{ZhaFenZhu03}; Feng, Leckman and Zhang, \citeyear{FenLecZha04}; Zhang, Liu and Wang, \citeyear{ZhaLiuWan10}), especially
in plant science (Rao and Xu, \citeyear{RaoXu98}; Xu and Xu, \citeyear{XuXu06}), insufficient
attention has been paid to addressing the unique challenges of
analyzing ordinal traits. Some researchers have recognized that it is
difficult to conduct genetic analyses of ordinal traits because such
traits cannot be directly characterized by a linear function of genetic
and environmental effects (Rao and Xu, \citeyear{RaoXu98}). To fill in this
methodological gap, we have made a systematic effort to develop
statistical methods for segregation analysis (Zhang, Feng and Zhu, \citeyear{ZhaFenZhu03}),
linkage analysis (Feng, Leckman and Zhang, \citeyear{FenLecZha04}) and association analysis (Zhang, Wang and Ye, \citeyear{ZhaWanYe06}) of ordinal traits (for family studies and case-control
studies).

%s2.1.1 ###
\subsubsection{Analysis of family data}\label{2010-03-16-1}

\begin{figure}
%%
%40){\circle{40}} \put(-78, 40){Mother} \put(-160, 40){\line(1,0){80}}
%-30){\circle{40}} \put(-168, -30){Sister} \put(-110, -50){\framebox(40,
%40)} \put(-109, -30){Brother} \put(-30, -30){\circle{40}} \put(-50,
%-30){\small Proband} \put(-10, -30){\line(1,0){40}} \put(30,
%-50){\framebox(40, 40)} \put(31, -30){Spouse} \put(10,
%-30){\line(0,-1){30}} \put(-10, -100){\framebox(40, 40)} \put(-5,
%-80){Son}
%

\includegraphics{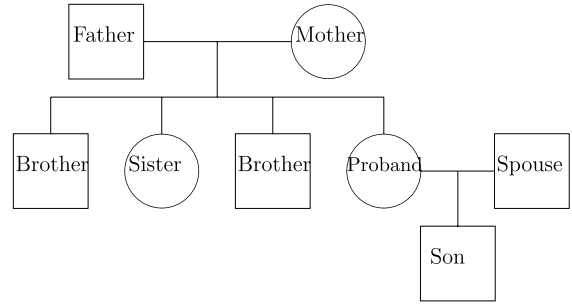}

\caption{A three-generation pedigree.} \label{2010-03-09-1}
\end{figure}

Long before the era of genomics, researchers collected data in
families, also called pedigrees as illustrated in Figure
\ref{2010-03-09-1}. Although the ascertainment process for families
varies, Figure \ref{2010-03-09-1} depicts a representative
three-generation pedigree. The proband is the first person who enters
into the study according to defined inclusion and exclusion criteria:
such criteria are related to the disease of interest. Other members of
the proband's family are included and directly or indirectly assessed,
depending on the circumstance. The key idea in analyzing family data
is that if a gene is a major driving force behind a disease, a trace in
the concordance of diseases in family members would reflect the
transmission pattern of a gene under the Mendelian laws. This is the
fundamental concept that Cannon and Rosanoff (\citeyear{CanRos11}) employed. This~ty\-pe
of analysis is referred to as segregation analysis.

%f1 ###

The Elston--Stewart (\citeyear{ElsSte71}) algorithm set up the quintessential
framework to analyze data from general pedigrees through a technique
called peeling. The main complication in analyzing pedigree data is the
complex relationship among family members, making it difficult to
express the likelihood function in an easily computed form. The peeling
algorithm makes use of the conditional independence embedded in the
pedigree resulting from the Mendelian laws, and so peels off the
complete likelihood function into smaller pieces before putting them
back together.

Other methods have been relatively recently developed using the concept
of latent random variables (Hopper, \citeyear{Hop89}; Babiker and Cuzick, \citeyear{BabCuz94}; Li
and Thompson, \citeyear{LiTho97}; Siegmund and McKnight, \citeyear{SieMcK98}; Zhang and Merikangas,
\citeyear{ZhaMer00}), which are closely related to the classic ousiotype models of
Cannings, Thompson and
  Skolnick (\citeyear{CanThoSko78}) in pedigree analysis. The basic idea is to use
latent variables to represent the contribution of unobserved factors
including a major gene, residual genetic factors and common
environmental factors. As discussed by Zhang and Merikangas (\citeyear{ZhaMer00}), the
computation involving pedigrees is similar to the peeling algorithm.
Advantages of using latent variable based models are that the
interactions between underlying genetic effects and the observed
covariates (e.g., demographic variables) can be considered.
Additionally, more relevant to this article, we can accommodate ordinal
traits in the latent variable framework.

%s2.1.2 ###
\subsubsection{A latent variable model}

We follow the notation of Zhang and Merikangas (\citeyear{ZhaMer00}) and Zhang, Feng and Zhu
(\citeyear{ZhaFenZhu03}). Consider a trait, $Y,$ that takes an ordinal value of $0, 1,
\ldots,K$. Let $\mathbf{x}$ be a $p$-vector of covariates that is also
available for each study subject. Three types of latent random
variables $U_1^i, U_2^i$ and $U_3^i$ are introduced within family $i$
to represent, respectively, (a)~common, unmeasured environmental
factors; (b)~genetic susceptibility of the family founders (a founder
refers to a subject whose parents are not a part of the observed
pedigree, e.g., father, mother and spouse in Figure~\ref{2010-03-09-1});
and (c)~the transmission of susceptibility genes from a parent to an
offspring.

The concept of latent variables is straightforward, but the interesting
and difficult part lies in the specification of their distributions.
They need to be interpretable and convenient. The following are the
assumptions that we found useful:

\begin{itemize}
\item$U_1^i$ follows Bernoulli distributions
$P\{U_1^i=1\}=1-P\{U_1^i=0\}=\theta_1,$ where $\theta_1$ is an unknown
parameter.
\item$U_2^i=(U_{2, 1}^i, U_{2, 2}^i, \ldots, U_{2, 2n_i-1}^i, U_{2,
2n_i}^i)',$ where $n_i$ is the size of pedigree $i.$ Here,
$P\{U_{2,2j-1}^i=1\}=1-P\{U_{2,2j}^i=0\}=\theta_2$ when $U_{2,2j-1}^i$
and $U_{2,2j}^i$ are the $U_2^i$-variables of a founder.
\item$U_3^i=(U_{3,1}^i, \ldots, U_{3, s_i}^i)'.$ According to the
Mendelian laws,\vspace*{1pt} $P\{U_{3, j}^i=1\}=P\{U_{3, j}^i=0\}=\frac{1}{2}, j=1,
\ldots, s_i,$ and $s_i$ is the number of parent-offspring pairs in
family $i.$ $U_3^i$ facilitates the transmission of $U_2^i$-variables
from the founders to the offspring. For example, if a parent of subject
$j$ has $U_2^i$-variables, $U_{2,2k-1}^i$ and $U_{2,2k}^i,$ and the
$U_3^i$-variable for this parent-offspring pair is $U_{3,l}^i,$ then
one of subject $j$'s $U_2^i$-variables is
$U_{2,2j-1}^i=U_{2,2k-1}^iU_{3,l}^i+U_{2,2k}^i(1-U_{3,l}^i).$
\item All latent variables are independent.
\end{itemize}

$U_1^i$ is a simple ``switch'' indicating the presence or absence of a
shared environment factor within family $i.$ $U_2^i$ is assigned
independently to each of founders who are the source for any gene to
enter into a family, and thus mimics the transmission of a~single major
susceptibility locus with alleles A and a of frequencies $\theta_2$ and
$1-\theta_2,$ respectively.

Conditional on all of the latent variables, denoted by $U^i$, within
family $i$, the probability distribution for member $j$ is assumed to
be
%
%e2.2 ###
%e2.1 ###
\begin{eqnarray}\label{2010-03-12-1}
&&\quad\hspace*{4pt} P\{Y_j^i\le k|U^i\}= \frac{\exp(\mathbf{x}_j^i {\bf\beta} +\alpha_k
+\mathbf{a}_j^i \mathbf{\gamma})} {1+ \exp(\mathbf{x}_j^i {\bf
\beta}
+\alpha_k +\mathbf{a}_j^i \mathbf{\gamma})},
\\
\eqntext{k=0, \ldots, K-1,}
\end{eqnarray}
where\vspace*{2pt} $\mathbf{a}_j^i=(U_1^i,U_{2,2j-1}^i+U_{2,2j}^i,
U_{2,2j-1}^iU_{2,2j}^i)^T$, and~$\mathbf{\beta}$ and $\mathbf{\gamma
}$ are $p$- and $3$-vectors of parameters. The $\alpha_k$ is the trait
level dependent intercept, $k=0,\ldots,K-1.$

As Zhang, Feng and Zhu (\citeyear{ZhaFenZhu03}) pointed out, the~$\beta$ parameters measure the
strength of association between the trait and the covariates,
conditional on the latent variables. The $\gamma$ parameters indicate
the familial and genetic contributions to the trait. The mode of
inheritance can be inferred from $\mathbf{\gamma}.$ For example,
$\gamma_2=0$ and $\gamma_3 \ne0$ suggests a recessive effect.

The likelihood function can be derived from (\ref{2010-03-12-1}). Due
to the presence of latent variables, the EM algorithm (Dempster, Laird
and Rubin, \citeyear{DemLaiRub77}) is the most convenient choice for parameter
estimation\break (Guo and Thompson, \citeyear{GuoTho92}; Zhang and Merikangas, \citeyear{ZhaMer00}; Zhang, Feng and Zhu, \citeyear{ZhaFenZhu03}). Although Zhang and Merikangas (\citeyear{ZhaMer00}) and Zhang, Feng and Zhu (\citeyear{ZhaFenZhu03})
presented an effective solution (e.g., a~modified likelihood), we
should note that the lack of concavity in the likelihood function makes
it a~challenging task to find the maximum likelihood estimates of the
model parameters. In addition, the $\theta$'s and $\gamma$'s are not
fully identifiable. The identifiability issue not only causes
computational problems, but also presents theoretical challenges in
statistical inference. Another important, yet understudied, issue is
the validation of the assumptions on the distributions of the latent
variables.

%%%%
But, how useful is the latent variable model (\ref{2010-03-12-1})?
First, it provides a regression framework to assess familial
aggregation and genetic contribution, and possibly interactions between
measured covariates and latent factors. Using data from a family study
of substance use (Merikangas et al., \citeyear{Meretal98}), Zhang and Merikangas (\citeyear{ZhaMer00})
were able to present extremely significant evidence of familial
aggregation $p$-value ${<}10^{-9}$ for alcohol dependence. This study
additionally demonstrated that transmission does not follow a major
locus pattern. In retrospect, their findings predicted the difficulty
of identifying major genes associated with alcoholism. In addition,
Zhang and Merikangas (\citeyear{ZhaMer00}) presented simulation examples to delineate
when the absence of latent variables in (\ref{2010-03-12-1}) affects
the estimates of the effects by the measured covariates. For example,
hypothetically, if the greater presence of females in a family has an
impact on the well-being of the family, ignoring the familial latent
variables is likely to result in a biased estimate of the sex
difference.

%t1 ###
\begin{table}
\caption{The probability estimates of rejecting $\gamma_1=0$ at the
significance level of 0.05. The covariate is omitted from the testing
despite the fact that its coefficient $\beta$ may not be zero}
\label{06142010-1}
\begin{tabular*}{\columnwidth}{@{\extracolsep{\fill}}lccc@{}}
\hline
& $\bolds{\gamma_1=0}$ & $\bolds{\gamma_1=1}$&$\bolds{\gamma_1=2}$\\
\hline
{$\beta=0$} & 0.0494 & 0.9503 & 1.0\\
{$\beta=1$} & 0.0534 & 0.9843 & 1.0\\
{$\beta=5$} & 0.1667 & 0.9971 & 1.0\\
{$\beta=10$} & 0.3828 & 0.9890 & 1.0\\
\hline
\end{tabular*}
\end{table}
%
%%%

Not only is it important to include the latent factors, but also it is
important to adjust for covariates. To further illustrate this point,
Zhang, Feng and Zhu (\citeyear{ZhaFenZhu03}) reported the following simulation. Ten thousand
data sets were generated from model (\ref{2010-03-12-1}) with
$\theta_1=0.3,$ $\beta$ chosen from 0, 1, 5 or 10, $\gamma_1$ from
0, 1
or 2, $\alpha_0=-1$ and $\alpha_1=1.$ To focus on the difference of
having or not having covariates, they set $\gamma_2=\gamma_3=0$. Each
data set consists of 200 families with 7 family members (similar to
Figure~\ref{2010-03-09-1}). One covariate $x$ was generated as follows.
For family $i$, $U_{1}^i$ were generated according to whether a random
number $r_{i1}$ from the $\operatorname{uniform}(0,1)$ was greater than 0.3 or not. For
member $j$ in family $i$, an independent random number $r_{ij2}$ from
the $\operatorname{uniform}(0,1)$ was generated. Then, $x_{ij}=0.9r_{ij2}+0.2r_{i1}.$

To evaluate the performance of the test statistic, the covariate was
deliberately ignored in the test. When $\beta=0,$ the covariate played
no role in the data generating process. The row corresponding to
$\beta=0$ in Table~\ref{06142010-1} displays the $p$-value (the column
corresponding to $\gamma_1=0$) and the power for two values of~%
$\gamma_1$ (1 or 2).

When $\beta\ne0,$ the covariate plays a role in the data generating
model. The data in Table~\ref{06142010-1} reveal the consequence of
ignoring the covariate, which is more severe when the effect of the
covariate is greater.

%s2.1.3 ###
\subsubsection{Linkage analysis}

While linkage analysis has a long history, it only became a common
practice after the availability of several convenient computing
programs (Ott, \citeyear{Ott74}; Kruglyak et al., \citeyear{Kruetal96}; Almasy and Blangero, \citeyear{AlmBla98}).
For statisticians, some of the common terminologies in linkage analysis
are puzzling, including the so-called LOD-score method and
nonparametric method.

Morton (\citeyear{MOR55}) first introduced the term ``LOD-score.'' LOD stands for
``the logarithm (base 10) of odds.'' The ``odds'' is a probability
ratio, or likelihood ratio, of the probability under an alternative
hypothesis to the probability under the null hypothesis. The LOD-score
method is essentially a~log-likelihood ratio test with two fundamental
differences: (a) the use of the base 10 logarithm versus the natural
logarithm; (b) the log-likelihood ratio statistic has a multiplier of 2
conforming to a $\chi^2$ distribution under certain regularity
conditions.

Specifically, the LOD-score is the log(base 10)-ratio of the likelihood
when the recombination fraction is less than $1/2$ (i.e., two loci are
not on the same chromosome, or called unlinked), to the likelihood when
the recombination fraction is $1/2$ (no linkage). The recombination
fraction is the frequency that a chromosomal crossover occurs between
two loci (or genes) during meiosis; 1\% of combination frequency is
termed the distance of one centimorgan (cM) in a genetic linkage map.
Because the LOD-score is in base 10, a score of 3 indicates 1000 to 1
odds in favor of the linkage, which is the conventional threshold for
declaring the evidence for linkage. If we convert a LOD-score of 3 into
the standard log-likelihood ratio statistic, it yields a $p$-value of
$2\times10^{-4}$ under $\chi_1^2.$ By Bonnferoni correction, it
corresponds to a genomewide $p$-value of 0.05 for 250 markers. This
number is in the range for the number of microsatellites used in
typical linkage studies.

In order to compute the LOD-score, we first need a~number of parameters
that determine the likelihood for a given recombination fraction. Then
use the maximum likelihood over the recombination fraction for the
likelihood under the alternative hypothesis. The parameters that are
required include the mode of inheritance, penetrance, and disease
allele frequency. These parameters are generally unknown and difficult
to estimate for complex diseases including mental illnesses. For
example, using segregation analysis (see Section~\ref{2010-03-16-1})
Pauls and Leckman (\citeyear{PauLec86}) examined specific genetic hypotheses about the
mode of transmission of Gilles de la Tourette's syndrome, by performing
segregation analyses in 30 nuclear families (two-generation pedigrees).
They\break concluded that Tourette's syndrome is inherited as an autosomal
dominant trait (one copy of the abnormal allele is sufficient to cause
the disease). The penetrance (the probability of having the disease for
a given genotype) was reported at 0.71 in males and 1.0 in females with
at least one abnormal allele. After several decades of research, no
major genetic variant has been identified for Tourette's syndrome, and
most likely this syndrome involves multiple genes, interacting with
environmental factors. This reality makes it difficult to infer the
mode of inheritance, penetrance, and disease allele frequency, and
conceptually, this may not make sense for complex diseases
(non-Mendelian inheritance).

This difficulty is somewhat alleviated since the LOD-score method has
been found to work reasonably well (e.g., Abreu, Greenberg and
Hodge, \citeyear{AbrGreHod99}) under
various parameter settings. There have been some efforts to improve the
robustness of the method (Gastwirth, \citeyear{Gas66,Gas85};
Whittemore, \citeyear{Whi96}). See\break
Zheng et al. (\citeyear{Zheetal09}) for a thorough review. Existing methods do not
extend to the case of ordinal traits. The effectiveness of the robust
methods remains to be studied. Naturally, nonparametric linkage methods
have been developed to avoid specification of the genetic model
parameters. In statistics, ``nonparametric'' methods typically refers
to distribution-free methods such as rank-based tests and methods based
on the empirical distribution. In linkage analysis, however,
``nonparametric'' does not mean ``distribution-free,'' but instead
refers to the replacement of true genetic model parameters with the
parameters of inheritance of markers, hypothesized to be close to the
disease locus. Thus, with nonparametric linkage methods, we still need
to compute the likelihood. Two core algorithms are used to compute the
likelihood: the Elston--Steward algorithm (\citeyear{ElsSte71}) and the Lander--Green
(\citeyear{LanGre87}) algorithm. As previously discussed, the Elston--Steward
algorithm (\citeyear{ElsSte71}) is a peeling algorithm that makes the computation in a
large pedigree feasible by splitting it into small pieces. This
algorithm was implemented in early versions of linkage analysis
programs (e.g., LIPED and LINKAGE); computational time increased
linearly in family size, but exponentially with the number of loci.
More recent programs (e.g., GENEHUNTER) use the Lander--Green (\citeyear{LanGre87})
algorithm that has first-order complexity in the number of loci, but
unfortunately exponential in the family size. Although Markov chain
Monte Carlo methods have been used to accommodate linkage analysis of
large families and a large number of markers (Guo and Thompson, \citeyear{GuoTho92}),
in practice, one may have to break large pedigrees apart in order to
run programs such as GENEHUNTER.

We should note that there had not been a linkage analysis program to
handle ordinal traits until the release of LOT (Zhang et al., \citeyear{Zhaetal08}).
Typically, the methods for linkage analysis can be divided into two
main steps; only the second step involves the trait (Kruglyak et al.,
\citeyear{Kruetal96}). The first step infers how genetic information travels in a
family as represented by the so-called ``inheritance vector.''

We will use the pedigree in Figure \ref{2010-03-09-1} to illustrate
this concept. The two parents and spouse are the founders of the
family, meaning that their parents are not in the current pedigree. The
four siblings and the child are nonfounders. The inheritance pattern at
marker locus $t$ is completely described by an inheritance vector $v(t)
= (v_{1}, v_{2},v_{3},v_{4}, \ldots, v_9, v_{10})'$. In other words, we
devote two elements for every nonfounder. The founders are not included
because they are the sources of the genes in the family and the
inheritance vector is conditional on their genes. The paired elements
describe the outcomes of the paternal and maternal meioses transmitted
to the nonfounders. Specifically, $v_{2j - 1} = 1$ or 2 according to
whether the grand paternal or grand maternal allele is transmitted in
the paternal meiosis to the $j$th nonfounder. $v_{2j}$ carries the
similar information for the corresponding maternal meiosis, namely,
$v_{2j}= 3$ or 4 according to whether the grand paternal or grand
maternal allele was transmitted in the maternal meiosis to the $j$th
nonfounder.

In practice, the genetic markers do not always allow us to determine
the true inheritance vector. In this case, the inheritance distribution
is the conditional probability distribution over the possible
inheritance vectors that conform with the alleles observed at $t,$
which we denote by $p{\{}v(t) = w{\}}$ for all inheritance vectors $w
\in V;$ here $V$ is the set of all possible inheritance vectors. In the
absence of any genotypic information, all inheritance vectors are
equally likely according to Mendel's first law; the probability
distribution is uniform.

For segregation analysis, we employed latent variables to reflect the
``imaginative'' genetic effects\break in~(\ref{2010-03-12-1}). In linkage
analysis, we have genetic markers that flow through the inheritance
vector. Thus, we can still use (\ref{2010-03-12-1}) for linkage
analysis except that~$\mathbf{a}_j^i$ should be
$(U_1^i,U_{2,v_{2j-1}}^i+U_{2,v_{2j}}^i).$ On one hand, we have a
reduced number of latent variables. On the other hand, many of the
latent variables depend on each other through the inheritance vectors.
The computation of the likelihood would be summed over all inheritance
vectors $w$ in $V,$ in addition to the probability space of the
remaining independent latent variables. Because of this connection and
distinction, the challenges in the linkage analysis of ordinal traits
are, to a great extent, similar to those in segregation analysis of
ordinal traits, for example, the asymptotic mixture of
$\chi^2$-distributions and the need to introduce the penalized
likelihood (Liang and Rathouz, \citeyear{LiaRat99}; Zhang, Feng and Zhu, \citeyear{ZhaFenZhu03}).

%s2.1.4 ###
\subsubsection{Association test}\label{2010-03-29-01}

As discussed above, linkage analysis focuses on testing the position of
a~marker, although it has been difficult to replicate findings in
linkage studies of mental disorders. An association analysis, however,
tests whether a genetic variant, including particular allele or
genotype of a marker and a haplotype in several markers, is associated~%
with a trait. Some study cohorts recruited for linkage~stu\-dies have
been re-genotyped for genomewide associa\-tion analyses. For binary or
quantitative traits, many methods have been developed and implemented.
Two commonly used programs are PLINK (Purcell et al., \citeyear{Puretal07}) and FBAT
(Rabinowitz and Laird, \citeyear{RabLai00}). To analyze an ordinal trait, Zhang, Wang and Ye
(\citeyear{ZhaWanYe06}) introduced the following proportional odds model:
%
%e2.3 ###
\begin{equation} \label{2010-03-31-1}
\logit\{P(y_{ij} \le k|G_{ij})\} = \alpha_{k} + \beta c_{ij},
\end{equation}
where $\alpha_{0}, \ldots, \alpha_{K-1}$ are non-descending level
parameters, $\beta$ is the genetic effect. The genetic factor $c_{ij}$
can be chosen to reflect the underlying mode of inheritance such as the
number of the risk allele. Under model (\ref{2010-03-31-1}), the null
hypothesis is $H_0\dvtx  \beta=0.$ The score statistic is
%
%e2.4 ###
\begin{equation} \label{2010-03-31-2}
S = \sum_{i,j} [R^+(y_{ij})-R^-(y_{ij})]A_{ij},
\end{equation}
where $R^+(y_{ij})$ and $R^-(y_{ij})$ are the counts of offspring in
the entire sample whose trait values are greater or less than $y_{ij}$,
respectively, and $A_{ij}$ is the number of copies of transmitted
alleles at the marker locus. Thus, Zhang, Wang and Ye
(\citeyear{ZhaWanYe06}) proposed the
following O-TDT test based on the score statistic:
\[
\frac{[S-E(S | Y)]^2}{\operatorname{Var}(S | Y)},
\]
which follows a $\chi_1^2$-distribution asymptotically. For a
case-control study, $R^+(y_{ij})$ and $R^-(y_{ij})$ are the numbers of
subjects whose trait values are greater or less than $y_{ij}$,
respectively.

If we rewrite the statistic in (\ref{2010-03-31-2}) in a general form
as $\sum_{i,j} w_{ij}A_{ij},$ this yields the classic TDT when
$w_{ij}=1$ and the QTDT (Rabinowitz, \citeyear{Rab}) when $w_{ij}=y_{ij}-{\bar
y},$ where $\bar y$ is the average of all $y_{ij}$'s. In other words,
all of these tests are a weighted function of the number of transmitted
alleles at the marker locus, and the choice of the weights depends on
the property of the trait. With this observation, after the proper
weights are computed, the existing FBAT software can be used to test
the association between any trait and alleles at a marker locus.

In the following, we describe a unified method to choose weights for
any kind of trait. It is straightforward to categorize a quantitative
trait into any reasonable number of categories (such as deciles) and
induce an ordinal scaled trait. This would allow the use of the O-TDT
for a quantitative trait. In their simulation studies, Zhang, Wang and Ye
(\citeyear{ZhaWanYe06}) demonstrated that this strategy has comparable power to the QTDT
for quantitative traits. This is due to the fact that the number of
categories is enough to capture most of the information in the data
(e.g., following Cochran's rule; Cochran, \citeyear{Coc77}). The advantage is that
the ordinal scaled test is not affected by the nonnormal distribution
of a~quantitative trait, and so, the unified approach is robust.

One limitation of the test proposed by Zhang, Wang and Ye
(\citeyear{ZhaWanYe06}) is that it
does not adjust for covariates. Environmental factors or covariates,
such as gender and age, may confound the association of interest. In a
subsequent work, Wang, Ye and Zhang (\citeyear{WanYeZha06}) generalized model
(\ref{2010-03-31-1}) to include covariates as follows:
%
%e2.5 ###
\begin{equation} \label{2010-03-31-3}
\hspace*{17pt}\logit\{P(y_{ij} \le k|G_{ij}), z_{ij}\} = \alpha_{k} + \beta c_{ij}+
\delta' z_{ij},\hspace*{-17pt}
\end{equation}
where $z_{ij}$ denotes the covariates and $\delta$ is the vector of the
corresponding coefficients. Consequently, the score statistic becomes
%
%e2.6 ###
\begin{equation} \label{2010-03-31-4}
S = \sum_{i,j} [\hat{\gamma}(y_{ij}, z_{ij})- \hat{\gamma}(y_{ij}-1,
z_{ij})]A_{ij},
\end{equation}
where
\[
\hat{\gamma}(k, z)=\frac{\exp(\hat{\alpha}_{k} + \hat{\delta}'
z_{ij}) }{1+\exp(\hat{\alpha}_{k} + \hat{\delta}' z_{ij})},
\]
which is
the estimated probability of having a trait value no greater than $k.$
Thus, the weight function in (\ref{2010-03-31-4}) is the difference
between the probability of having a trait value greater than $y_{ij}$
and the probability of having a trait value less than $y_{ij}.$ Not
surprisingly, this is in essence the same as the weight function in
(\ref{2010-03-31-2}) where we used counts instead of frequency (or
probability).

It is important to note that association analysis does not directly
equate to a causal relationship. In well-designed genetic association
studies, an observed association is expected to result from either a
causal functional variant of a gene, or the linkage disequilibrium
between the marker and a susceptibility gene. In population-based
case-control studies, there are typically attempts to match cases and
controls by important demographic and/or baseline information. It is
not wise to over-match subjects. Alternatively, we can collect
potentially important environmental variables and consider them in the
association analysis. We can also use principal component analysis on
the genotypes to explore whether there are ``clusters'' in the study
cohorts that are not appropriately reflected in the environmental
variables. In family-based studies, the association tends to be
conditional on parental genotypes and all phenotypes.

%s2.1.5 ###
\subsubsection{Unique challenges in analyzing ordinal traits}

Understanding the genetic mechanisms for complex diseases is
challenging regardless of whether we analyze binary, ordinal or
continuous traits. Any challenges that exist for analyzing binary and
continuous traits remain for ordinal traits. What are the unique
challenges in analyzing ordinal traits? The key difference is that
there is not a simple distribution function for ordinal traits. For
continuous traits, the assumption is that the traits can somehow be
treated under normality, by transformation if needed. For binary
traits, through a link function (e.g., logit) we only need to deal with
a Bernoulli distribution. However, for ordinal traits, the two typical
approaches are (a) to assume a reliability variable or a continuous
latent variable or (b) to assume a proportional odds model as we
presented above. The first challenge is in the estimation. The
likelihood function is complicated, and based on the numerical results,
it has multiple local maxima. In addition, due to identifiability (or
near-identifiability), the likelihood function may be relatively flat.
Combinations of the EM and other algorithms can provide practical
solutions, but finding a more efficient algorithm is an open problem.

The second challenge is in the inference. When latent variables or
mixture distributions are used, some of the commonly assumed regularity
conditions do not hold. One solution is to use a penalized likelihood
function (Zhang, Feng and Zhu, \citeyear{ZhaFenZhu03}) that prevents the parameters from being
near the singularity points.

Finally, model diagnostics are difficult. For example, how do we know
the latent variable-based model or the proportional odds model provides
an adequate fit to the data? Although the models and methods presented
above do not address this and other questions, they provide a
foundation for further research and improvement.

%s2.2 ###
\subsection{Comorbidity}

The methods described above only deal with a~sing\-le trait. However,
comorbidity is the rule rather~than the exception in studies of mental
and behavioral disorders. For example, a patient may suffer from both
anxiety and depression (Li and Burmeister, \citeyear{LiBur09}), and the same patient
may also be addicted to nicotine, alcohol, or other substances
(Me\-rikangas et al., \citeyear{Meretal98}; True et al., \citeyear{Truetal99}). From a data analysis
perspective, we need to consider how important it is to accommodate
multiple diseases/traits. In a real-data example, Chen et al. (\citeyear{Cheetal11})
analyzed a data set from the Study of Addiction: Genetics and
Environment (SAGE). By simply considering addiction to at least two of
the six substances (addiction to nicotine, alcohol, marijuana, cocaine,
opiates or other drugs), we were able to identify the PKNOX2 gene that
reached genomewide significance level among European-origin females.
Interestingly, the PKNOX2 gene has been previously identified as one of
the cis-regulated genes for alcohol addiction in mice (Mulligan et al.,
\citeyear{Muletal06}). To further delineate the benefit of considering multivariate
traits, Zhu and Zhang (\citeyear{ZhuZha09}) conducted comprehensive simulation
studies, considered the correlations of 0.2 and $-0.2$ among three
quantitative traits, and demonstrated that tes\-ting correlated traits
jointly is more powerful than testing a single trait at a time. Using
generalized estimation equation, Lange et al. (\citeyear{Lanetal03}) developed
a~family-based association test for multivariate quantitative traits
(FBAT-GEE). Recently, Zhang, Liu and Wang (\citeyear{ZhaLiuWan10}) constructed a nonparametric
test based on the generalized Kendall's tau to accommodate any
combination of dichotomous, ordinal, and quantitative traits.

%s2.2.1 ###
\subsubsection{Kendall's tau}

Kendall's $\tau$ is a rank-based correlation between two variables. It
contracts the probability of observing the two variables in the~sa\-me
order in two observations with the probability~of observing the two
variables in the opposite order. Specifically, for a sample of $n$
observations $(X_1, Y_1),\break \ldots, (X_n, Y_n),$ two observations $(X_i,
Y_i)$ and $(X_j, Y_j)$ are called concordant if $(X_i- X_j)(Y_i-Y_j)>0$
and discordant if $(X_i-X_j)(Y_i-Y_j)<0.$ Then Kendall's~$\tau$ is
based on the difference between the numbers of concordant pairs and
discordant pairs.

We introduce a kernel function,
\begin{eqnarray*}
&&
\phi( (X_i, Y_i), (X_j, Y_j) )\\
&&\quad=  \operatorname{sign}\{ (X_i- X_j)(Y_i-Y_j)\}
\\
&&\quad=
\cases{
1, &if $(X_i- X_j)(Y_i-Y_j)>0$, \cr
-1, &if $(X_i- X_j)(Y_i-Y_j) < 0$,\cr
0, &if $(X_i- X_j)(Y_i-Y_j)=0$,}
\end{eqnarray*}
and define a $U$-statistic\vspace*{-2pt}
%
%e2.7 ###
\begin{equation}\label{112408-1}
U =\pmatrix{n\cr2}^{-1} \sum_{i<j} \phi( (X_i, Y_i), (X_j, Y_j) ).\vspace*{-2pt}
\end{equation}
Then, Kendall's $\tau$ is\vspace*{-2pt}
%
%e2.8 ###
\begin{equation}\label{112408-2}
\tau=\frac{U}{\sqrt{ \operatorname{Var}_0 (U)}},\vspace*{-2pt}
\end{equation}
where $\operatorname{Var}_0 (U)$ is the variance of $U$ under the null
hypothesis of no correlation between $X$ and $Y,$ and equal to
$n(n-1)(2n+5)/18$ if $X$ and $Y$ are continuous variables (Hollander
and Wolfe, \citeyear{HolWol99}).

%s2.2.2 ###
\subsubsection{Generalized Kendall's tau}

To test the association between genetic markers and comorbidity, Zhang, Liu and Wang (\citeyear{ZhaLiuWan10}) generalized Kendall's tau as follows. For individuals $i$
and $j$, let $T_i$ and $T_j$ be their vectors of traits, respectively.
Then, a trait kernel is defined as
\[
F_{ij} = \bigl(f_1\bigl(T_i^{(1)} - T_j^{(1)}\bigr), \ldots, f_p\bigl(T_i^{(p)} -
T_j^{(p)}\bigr)\bigr)',
\]
where function $f_k(\cdot)$ is the identity function for\break a~quantitative
or binary trait (Rabinowitz, \citeyear{Rab}), or the sign function for an ordinal
trait (Zhang, Wang and Ye, \citeyear{ZhaWanYe06}).

Also, recall that, as in Section~\ref{2010-03-29-01}, $c$ is the number
of any chosen allele for marker genotype and let~$C_i$ refer to the $C$
for the $i$th subject. Then, Zhang, Liu and Wang (\citeyear{ZhaLiuWan10}) defined a marker
kernel as\vspace*{-2pt}
\[
D_{ij} = c_i - c_j.
\]
Their $U$-statistic is defined as
%
%e2.9 ###
\begin{equation}\label{2010-04-04-1}
U = \pmatrix{n \cr2}^{-1} \sum_{i<j} D_{ij} F_{ij}.
\end{equation}
The association test statistic, or generalized Ken\-dall's tau, is $U'
\var_0^{-1}(U) U$, where $\var_0(U)$ is the variance of $U$ under the
null hypothesis that there is no association between marker alleles and
any linked locus that influences the trait $T$. The test statistic
follows an asymptotic $\chi^2$-distribution under the null hypothesis.

Obviously, the statistic in (\ref{2010-04-04-1}) does not incorporate
covariate effects. This is relatively straightforward for a single
trait as was done in (\ref{2010-03-31-3}). Here, the traits can be a
hybrid of different traits. An alternative is to impose different
weights for each pair of samples in the statistic (\ref{2010-04-04-1})
according to the information of their covariates. The weight, denoted
by $w(z_i, z_j)$ for the pair $(i, j)$, reflects the relative
importance attributed by the covariates when we derive the statistic.
Zhu, Jiang and Zhang (\citeyear{ZhuJiaZha}) examined the following weight function. Write $z =
(z^{\mathrm{co}}, z^{\mathrm{ca}})'$ with $z^{\mathrm{co}} = (z^{(1)},\ldots,z^{(l_1)})'$ for the
continuous covariates and $z^{\mathrm{ca}} = (z^{(l_1+1)},\ldots,z^{(l)})'$ for
the categorical covariates. They defined the weight function $w(z_i,
z_j)$ as
%
%e2.10 ###
\begin{equation}\label{06102010-1}
\quad w(z_i, z_j) = W(\|z^{\mathrm{co}}_{i} - z^{\mathrm{co}}_{j}\|) I(z^{\mathrm{ca}}_{i} =
z^{\mathrm{ca}}_{j}),
\end{equation}
where $W(\cdot)$ is a positive and decreasing function, for example,
$W(u) = \exp(- u^2 / 2h^2)$, and $I(\cdot)$ is the indicator function.
Then a weighted test statistic is given by
%
%e2.11 ###
\begin{eqnarray}\label{2010-04-06-1}
S = \pmatrix{n \cr2}^{-1} \sum_{i<j} D_{ij} F_{ij} w(z_i, z_j).
\end{eqnarray}

Zhang, Liu and Wang (\citeyear{ZhaLiuWan10}) and Zhu, Jiang and Zhang (\citeyear{ZhuJiaZha}) showed that under the null
hypothesis, the test statistic $S$ (weighted or not) has the following
asymptotic distribution conditional on all phenotypes and parental
genotypes:
\[
\var^{-1/2}_0(S) [S - E_0(S)] \conD N(0, I_p),
\]
where
\begin{eqnarray*}
E_0(S) &=& \frac{2}{n-1} \sum_{i=1}^n \bar{u}_i E_0(C_i | M_i^{\mathrm{pa}}),
\\
\var_0(S) &=& \frac{4}{(n-1)^2} \\
&&{}\cdot\sum_{i=1}^n \sum_{j=1}^n \bar{u}_i
\bar{u}'_j \cov_0(C_i,C_j | M_i^{\mathrm{pa}},M_j^{\mathrm{pa}}).
\end{eqnarray*}
Consequently, the following test statistic
\[
\chi^2_{\mathrm{tau}}=[S - E_0(S)]' \var^{-1}_0(S) [S - E_0(S)]
\]
converges to $\chi^2_p$ in distribution under the null hypothesis
provided that $\var_0(S)$ is full rank. In a~ca\-se-control study, we do
not have the markers from parents and hence the conditional
expectations are replaced with the unconditional ones. Thus, the key
difference in the test statistics between family studies and population
studies lies in the conditioning on the parental markers. The
conditioning on the parental markers gives the family studies a major
advantage in removing the effect of population admixture, but family
studies tend to be more difficult and expensive to carry out.

Under the alternative hypothesis, the test statistic $\chi^2_{\mathrm{tau}}$ can
be written as a weighted sum of noncentral $\chi_1^2=\sum_{i=1}^p e_i
\chi^2_1(\phi_i),$ where $e_1 \ge\cdots\ge e_p$ are the nonnegative
eigenvalues of $\bSigma^{1/2}_1 \bSigma^{-1}_0 \bSigma^{1/2}_1$.
$\phi_i = \mu_{R_i}^2$ and~$\mu_{R_i}$ is the $i$th component of
$\bmu_R = Q \bSigma^{-1/2}_1 \bmu$, whe\-re $Q$ is an orthonormal matrix
such that\break $Q \bSigma^{1/2}_1 \bSigma^{-1}_0 \bSigma^{1/2}_1 Q' =
\diag
(e_1, \ldots, e_p)$. $\bmu$ is the difference\vadjust{\eject} in the means of $S$ under
the alternative and null hypotheses. Using the approximation theory of
Pearson (\citeyear{Pea59}), Solomon and Stephens (\citeyear{SolSte77}) and Liu, Tang and Zhang (\citeyear{LiuTanZha09}), we
can find a certain degree of freedom $l$ and noncentral parametric
$\upsilon$ such that the distribution of $\chi^2_{\mathrm{tau}}$ can be closely
approximated by $\chi^2_l(\upsilon).$ Through simulation studies, Zhu
et al. confirmed that this approximation is accurate enough for power
calculation.

It is noteworthy that the weight function in (\ref{06102010-1}) is
restrictive with respect to categorical covariates, especially so for
ordinal covariates. The use of genomic propensity score can give rise
to an alternative weight function. Specifically, for a di-allelic
marker $G$ (e.g., SNP), the genomic propensity score is the conditional
probability $p_g(z) \!=\! P(G\!=\! g | Z \!=\! z)$. This probability can be fitted
by a logistic regression model or proportional odds model depending on
whether $G$ is chosen as an allele type or genotype. In the latter
choice, the model also depends on the mode of inheritance. In the
current genomewide association studies, we usually only have genotypes
and cannot distinguish the phases of individual alleles. Thus, we have
to construct genomic propensity scores by considering various modes of
inheritance. Once the genomic propensity score is estimated, it can be
treated as a numerical covariate and then we can use (\ref{06102010-1})
again.

%s2.2.3 ###
\subsubsection{Examples}

Zhang, Liu and Wang (\citeyear{ZhaLiuWan10}) reanalyzed a data set from the Collaborative Study
on the Genetics of Alcoholism (COGA) (\cite{Beg95}; \cite{Edeetal05}). The data came from a multi-center
(9 sites) consortium that recruited study participants by requiring
every proband to meet two alcohol dependence diagnostic criteria ba\-sed
on DSM-IV-R (American Psychiatric Association, \citeyear{autokey5}). The first-degree
relatives of the probands were invited into the study. Zhang, Liu and Wang (\citeyear{ZhaLiuWan10}) included a total of 1614 individuals from 143 families. They
considered three phenotypes: (1) alcohol DX-DSM3R${}+{}$Feighner; (2) maximum
number of drinks in a 24-hour period; and (3) the response to ``spent
so much time drinking, had little time for anything else.'' Using the
first phenotype alone, the $p$-value of the association between a peak
marker D7S679 on chromosome 7 and the trait was 0.0019. However, when
the three traits are analyzed together, D7S679 remains the peak marker,
and the $p$-value is reduced to 0.00055, demonstrating the possibility
that the other two phenotypes enhanced the association signal. If the
other two phenotypes are analyzed alone, the analysis did not lead to
anything worthy of further attention.\vadjust{\eject}

In the analysis cited above, the association was assessed without
considering covariates. In a follow-up analysis, Zhu, Jiang and Zhang (\citeyear{ZhuJiaZha})
considered two important covariates: age at interview and sex. When
these two covariates were controlled for, the $p$-value of the
association between the peak marker D7S679 and the three phenotypes
went down further to 0.000313.

%s3 ###
\section{Discussion}

Studying comorbidity is a significant issue in mental and behavioral
research, dating back to a century ago (Cannon and Rosanoff, \citeyear{CanRos11}).
This is challenging due to a lack of statistical methods that
accommodate the complexity of comorbidity. While dealing with
comorbidity in genetic studies is the focus of this review, it is
achieved through gradual development, and accumulation of methods.
Various challenges are dealt with along the way.

Although I focused on the analysis of ordinal traits and applications
in mental health, the presented me\-thods are closely related to robust
and rank-based methods for binary and quantitative traits. Furthermore,
ordinal traits arise in studies of diseases besides mental illnesses,
such as cancer (specifically, different stages).

From the statistical perspective, the methods that are presented here
have broad applications beyond genetic association studies. From
college admissions, to job searches, to scientific investigations, we
make inferences based on multidimensional data. It is important and
imperative to consider and develop inferential tools for multivariate
outcomes, particularly when the outcomes are discrete. There is
extensive literature on the statistical analysis of multivariate normal
variables as well as on nonparametric tests for a single variable of
nonnormal distribution. However, few options are available for the
inference when we have multiple nonnormally distributed variables and
potential hybrids of continuous and discrete variables. To overcome
this challenge, I presented several useful statistical techniques such
as the rank-based $U$-statistics and the kernel-based\break weighted statistics
to accommodate the mix of continuous and discrete outcomes and the
presence of important covariates.

\section*{Acknowledgments}

This work is supported in part by National
Institute on Drug Abuse Grant R01DA016750. The author wishes to thank Professor David
Madigan for his encouragement on this review. He also thanks Dr. Gang
Zheng and Professor Joseph Gastwirth for helpful discussions and
comments, and Jennnifer Brennan for careful reading and comments.

% imsref loaded by arune.pranskunaite, 2011-04-11 13:20:30
%


\begin{thebibliography}{88}
% BibTex style file: ims.bst, 2010-03-23
% Default style options (sort=0,type=number).
% Used options (sort=1,type=nameyear).

%b1 ###
\bibitem[\protect\citeauthoryear{Abelson et al.}{2005}]{Abeetal05}
%
\begin{barticle}[pbm]
\bauthor{\bsnm{Abelson},~\bfnm{Jesse~F.}\binits{J.~F.}},
\bauthor{\bsnm{Kwan},~\bfnm{Kenneth~Y.}\binits{K.~Y.}},
\bauthor{\bsnm{O'Roak},~\bfnm{Brian~J.}\binits{B.~J.}},
\bauthor{\bsnm{Baek},~\bfnm{Danielle~Y.}\binits{D.~Y.}},
\bauthor{\bsnm{Stillman},~\bfnm{Althea~A.}\binits{A.~A.}},
\bauthor{\bsnm{Morgan},~\bfnm{Thomas~M.}\binits{T.~M.}},
\bauthor{\bsnm{Mathews},~\bfnm{Carol~A.}\binits{C.~A.}},
\bauthor{\bsnm{Pauls},~\bfnm{David~L.}\binits{D.~L.}},
\bauthor{\bsnm{Rasin},~\bfnm{Mladen-Roko}\binits{M.-R.}},
\bauthor{\bsnm{Gunel},~\bfnm{Murat}\binits{M.}},
\bauthor{\bsnm{Davis},~\bfnm{Nicole~R.}\binits{N.~R.}},
\bauthor{\bsnm{Ercan-Sencicek},~\bfnm{A.~Gulhan}\binits{A.~G.}},
\bauthor{\bsnm{Guez},~\bfnm{Danielle~H.}\binits{D.~H.}},
\bauthor{\bsnm{Spertus},~\bfnm{John~A.}\binits{J.~A.}},
\bauthor{\bsnm{Leckman},~\bfnm{James~F.}\binits{J.~F.}}, \bauthor
{\bsnm{Leon
S.~Dure},~\bfnm{4th}\binits{t.}},
\bauthor{\bsnm{Kurlan},~\bfnm{Roger}\binits{R.}},
\bauthor{\bsnm{Singer},~\bfnm{Harvey~S.}\binits{H.~S.}},
\bauthor{\bsnm{Gilbert},~\bfnm{Donald~L.}\binits{D.~L.}},
\bauthor{\bsnm{Farhi},~\bfnm{Anita}\binits{A.}},
\bauthor{\bsnm{Louvi},~\bfnm{Angeliki}\binits{A.}},
\bauthor{\bsnm{Lifton},~\bfnm{Richard~P.}\binits{R.~P.}},
\bauthor{\bsnm{Sestan},~\bfnm{Nenad}\binits{N.}} \AND
\bauthor{\bsnm{State},~\bfnm{Matthew~W.}\binits{M.~W.}}
(\byear{2005}).
\btitle{Sequence variants in SLITRK1 are associated with Tourette's syndrome}.
\bjournal{Science}
\bvolume{310}
\bpages{317--320}.
\bid{doi={10.1126/science.1116502}, issn={1095-9203}, pii={310/5746/317},
pmid={16224024}}
\end{barticle}
%
\endbibitem

%b2 ###
\bibitem[\protect\citeauthoryear{Abreu, Greenberg and
Hodge}{1999}]{AbrGreHod99}
%
\begin{barticle}[auto:STB|2011-03-03|12:04:44]
\bauthor{\bsnm{Abreu},~\bfnm{P.~C.}\binits{P.~C.}},
\bauthor{\bsnm{Greenberg},~\bfnm{D.~A.}\binits{D.~A.}} \AND
\bauthor{\bsnm{Hodge},~\bfnm{S.~E.}\binits{S.~E.}}
(\byear{1999}).
\btitle{Direct power comparisons between simple lod scores and NPL
scores for
linkage analysis in complex diseases}.
\bjournal{Am. J. Hum. Genet.}
\bvolume{65}
\bpages{847--857}.
\end{barticle}
%
\endbibitem

%b3 ###
\bibitem[\protect\citeauthoryear{Allison}{1997}]{All97}
%
\begin{barticle}[pbm]
\bauthor{\bsnm{Allison},~\bfnm{D.~B.}\binits{D.~B.}}
(\byear{1997}).
\btitle{Transmission-disequilibrium tests for quantitative traits}.
\bjournal{Am. J. Hum. Genet.}
\bvolume{60}
\bpages{676--690}.
\bid{issn={0002-9297}, pmcid={1712500}, pmid={9042929}}
\end{barticle}
%
\endbibitem

%b4 ###
\bibitem[\protect\citeauthoryear{Almasy and Blangero}{1998}]{AlmBla98}
%
\begin{barticle}[auto:STB|2011-03-03|12:04:44]
\bauthor{\bsnm{Almasy},~\bfnm{L.}\binits{L.}} \AND
\bauthor{\bsnm{Blangero},~\bfnm{J.}\binits{J.}}
(\byear{1998}).
\btitle{Multipoint~quantitati\-ve-trait linkage analysis in general pedigrees}.
\bjournal{Am. J. Hum. Genet.}
\bvolume{62}
\bpages{1198--1121}.
\end{barticle}
%
\endbibitem

%b5 ###
\bibitem[\protect\citeauthoryear{}{1994}]{autokey5}
%
\begin{bmisc}[auto:STB|2011-03-03|12:04:44]
\bauthor{\borganization{American Psychiatric Association}}
(\byear{1994}).
\bhowpublished{\textit{Diagnostic and
Statistical Manual of Mental Disorders}, 4th ed.
American Psychiatric Association Press, Washington, DC}.
\end{bmisc}
%
\endbibitem

%b6 ###
\bibitem[\protect\citeauthoryear{Amos}{1994}]{Amo94}
%
\begin{barticle}[auto:STB|2011-03-03|12:04:44]
\bauthor{\bsnm{Amos},~\bfnm{C.~I.}\binits{C.~I.}}
(\byear{1994}).
\btitle{Robust variance-components approach for assess-ing genetic
linkage in
pedigrees}.
\bjournal{Am. J. Hum. Genet.}
\bvolume{54}
\bpages{535--543}.
\end{barticle}
%
\endbibitem

%b7 ###
\bibitem[\protect\citeauthoryear{Arking et al.}{2006}]{Arketal06}
%
\begin{barticle}[pbm]
\bauthor{\bsnm{Arking},~\bfnm{Dan~E.}\binits{D.~E.}},
\bauthor{\bsnm{Pfeufer},~\bfnm{Arne}\binits{A.}},
\bauthor{\bsnm{Post},~\bfnm{Wendy}\binits{W.}},
\bauthor{\bsnm{Kao},~\bfnm{W~H~Linda}\binits{W.~H.~L.}},
\bauthor{\bsnm{Newton-Cheh},~\bfnm{Christopher}\binits{C.}},
\bauthor{\bsnm{Ikeda},~\bfnm{Morna}\binits{M.}},
\bauthor{\bsnm{West},~\bfnm{Kristen}\binits{K.}},
\bauthor{\bsnm{Kashuk},~\bfnm{Carl}\binits{C.}},
\bauthor{\bsnm{Akyol},~\bfnm{Mahmut}\binits{M.}},
\bauthor{\bsnm{Perz},~\bfnm{Siegfried}\binits{S.}},
\bauthor{\bsnm{Jalilzadeh},~\bfnm{Shapour}\binits{S.}},
\bauthor{\bsnm{Illig},~\bfnm{Thomas}\binits{T.}},
\bauthor{\bsnm{Gie\-ger},~\bfnm{Christian}\binits{C.}},
\bauthor{\bsnm{Guo},~\bfnm{Chao-Yu}\binits{C.-Y.}},
\bauthor{\bsnm{Larson},~\bfnm{Martin~G.}\binits{M.~G.}},
\bauthor{\bsnm{Wichmann},~\bfnm{H.~Erich}\binits{H.~E.}},
\bauthor{\bsnm{Marb{\'{a}}n},~\bfnm{Eduardo}\binits{E.}},
\bauthor{\bsnm{O'Donnell},~\bfnm{Christopher~J.}\binits{C.~J.}},
\bauthor{\bsnm{Hirschhorn},~\bfnm{Joel~N.}\binits{J.~N.}},
\bauthor{\bsnm{K{\"{a}}{\"{a}}b},~\bfnm{Stefan}\binits{S.}},
\bauthor{\bsnm{Spooner},~\bfnm{Peter~M.}\binits{P.~M.}},
\bauthor{\bsnm{Meitinger},~\bfnm{Thomas}\binits{T.}} \AND
\bauthor{\bsnm{Chakra\-varti},~\bfnm{Aravinda}\binits{A.}}
(\byear{2006}).
\btitle{A common genetic variant in the NOS1 regulator NOS1AP
modulates cardiac
repolarization}.
\bjournal{Nat. Genet.}
\bvolume{38}
\bpages{644--651}.
\bid{doi={10.1038/ng1790}, issn={1061-4036}, pii={ng1790}, pmid={16648850}}
\end{barticle}
%
\endbibitem

\bibitem[\protect\citeauthoryear{Babiker and Cuzick}{1994}]{BabCuz94}
\begin{barticle}[pbm]
\bauthor{\bsnm{Babiker},~\bfnm{A.}\binits{A.}} \AND
  \bauthor{\bsnm{Cuzick},~\bfnm{J.}\binits{J.}}
(\byear{1994}).
\btitle{A simple frailty model for family studies with covariates}.
\bjournal{Stat. Med.}
\bvolume{13}
\bpages{1679--1692}.
\bid{issn={0277-6715}, pmid={7973243}}
\end{barticle}
\endbibitem

%b8 ###
\bibitem[\protect\citeauthoryear{Begleiter}{1995}]{Beg95}
%
\begin{bmisc}[auto:STB|2011-03-03|12:04:44]
\bauthor{\bsnm{Begleiter},~\bfnm{H.}\binits{H.}}
(\byear{1995}).
\bhowpublished{The collaborative study on the genetics of alcoholism.
\textit{Alcohol Health Res. World}}
\bvolume{19}
\bpages{228--236}.
\end{bmisc}
%
\endbibitem

%b9 ###
\bibitem[\protect\citeauthoryear{Blackwelder and Elston}{1985}]{BlaEls85}
%
\begin{barticle}[pbm]
\bauthor{\bsnm{Blackwelder},~\bfnm{W.~C.}\binits{W.~C.}} \AND
\bauthor{\bsnm{Elston},~\bfnm{R.~C.}\binits{R.~C.}}
(\byear{1985}).
\btitle{A comparison of sib-pair linkage tests for disease susceptibility
loci}.
\bjournal{Genet. Epidemiol.}
\bvolume{2}
\bpages{85--97}.
\bid{doi={10.1002/gepi.1370020109}, issn={0741-0395}, pmid={3863778}}
\end{barticle}
%
\endbibitem

%b10 ###
\bibitem[\protect\citeauthoryear{Blangero and Almasy}{1997}]{BlaAlm97}
%
\begin{barticle}[pbm]
\bauthor{\bsnm{Blangero},~\bfnm{J.}\binits{J.}} \AND
\bauthor{\bsnm{Almasy},~\bfnm{L.}\binits{L.}}
(\byear{1997}).
\btitle{Multipoint oligogenic linkage analysis of quantitative traits}.
\bjournal{Genet. Epidemiol.}
\bvolume{14}
\bpages{959--964}.
\bid{doi={10.1002/(SICI)1098-2272(1997)14:6<959::AID-GEPI66>3.0.CO;2-K},
issn={0741-0395},
pii={10.1002/(SICI)1098-2272(1997)14:6<959::AID-GEPI66>3.0.CO;2-K},
pmid={9433607}}
\end{barticle}
%
\endbibitem

\bibitem[\protect\citeauthoryear{Cannings, Thompson and
  Skolnick}{1978}]{CanThoSko78}
\begin{barticle}[mr]
\bauthor{\bsnm{Cannings},~\bfnm{C.}\binits{C.}},
  \bauthor{\bsnm{Thompson},~\bfnm{E.~A.}\binits{E.~A.}} \AND
  \bauthor{\bsnm{Skolnick},~\bfnm{M.~H.}\binits{M.~H.}}
(\byear{1978}).
\btitle{Probability functions on complex pedigrees}.
\bjournal{Adv. Appl. Probab.}
\bvolume{10}
\bpages{26--61}.
\bid{issn={0001-8678}, mr={0490038}}
\end{barticle}
\endbibitem

%b11 ###
\bibitem[\protect\citeauthoryear{Cannon and Rosanoff}{1911}]{CanRos11}
%
\begin{bmisc}[auto:STB|2011-03-03|12:04:44]
\bauthor{\bsnm{Cannon},~\bfnm{G.~L.}\binits{G.~L.}} \AND
\bauthor{\bsnm{Rosanoff},~\bfnm{A.~J.}\binits{A.~J.}}
(\byear{1911}).
\bhowpublished{Preliminary report of a study of heredity in insanity in the
light of
the Mendelian laws. Reprinted from \textit{J. Nervous and Mental
Disorders}}
\bvolume{38}
\bpages{272--279}.
\end{bmisc}
%
\endbibitem

%b12 ###
\bibitem[\protect\citeauthoryear{Carter and Chung}{1980}]{CarChu80}
%
\begin{barticle}[auto:STB|2011-03-03|12:04:44]
\bauthor{\bsnm{Carter},~\bfnm{C.~L.}\binits{C.~L.}} \AND
\bauthor{\bsnm{Chung},~\bfnm{C.~S.}\binits{C.~S.}}
(\byear{1980}).
\btitle{Segregation analysis of schizophrenia under a mixed genetic model}.
\bjournal{Hum. Hered.}
\bvolume{30}
\bpages{350--356}.
\end{barticle}
%
\endbibitem

%b13 ###
\bibitem[\protect\citeauthoryear{Chen et al.}{2007}]{Cheetal07}
%
\begin{barticle}[auto:STB|2011-03-03|12:04:44]
\bauthor{\bsnm{Chen},~\bfnm{X.}\binits{X.}},
\bauthor{\bsnm{Liu},~\bfnm{C.~T.}\binits{C.~T.}},
\bauthor{\bsnm{Zhang},~\bfnm{M.~Z.}\binits{M.~Z.}} \AND
\bauthor{\bsnm{Zhang},~\bfnm{H.~P.}\binits{H.~P.}}
(\byear{2007}).
\btitle{A forest-based approach to identifying gene and gene--gene
interactions}.
\bjournal{Proc. Natl. Acad. Sci. USA}
\bvolume{104}
\bpages{19199--19203}.
\end{barticle}
%
\endbibitem

%b14 ###
\bibitem[\protect\citeauthoryear{Chen et al.}{2011}]{Cheetal11}
%
\begin{barticle}[auto:STB|2011-03-03|12:04:44]
\bauthor{\bsnm{Chen},~\bfnm{X.}\binits{X.}},
\bauthor{\bsnm{Cho},~\bfnm{K.}\binits{K.}},
\bauthor{\bsnm{Singer},~\bfnm{B.~H.}\binits{B.~H.}} \AND
\bauthor{\bsnm{Zhang},~\bfnm{H.~P.}\binits{H.~P.}}
(\byear{2011}).
\btitle{The nuclear transcription factor PKNOX2 is a candidate gene for
substance dependence in European-origin women}.
\bjournal{PLoS ONE}
\bvolume{6}
\bpages{e16002}.
\end{barticle}
%
\endbibitem

%b15 ###
\bibitem[\protect\citeauthoryear{Cochran}{1977}]{Coc77}
%
\begin{bbook}[mr]
\bauthor{\bsnm{Cochran},~\bfnm{William~G.}\binits{W.~G.}}
(\byear{1977}).
\btitle{Sampling Techniques}, \bedition{3rd} ed.
\bpublisher{Wiley, New York}.
\bid{mr={0474575}}
\end{bbook}
%
\endbibitem

%%b16 ###
%%
%(\byear{2008}).
%aid in
%gene identification}.
%%

\bibitem[\protect\citeauthoryear{Dempster, Laird and Rubin}{1977}]{DemLaiRub77}
\begin{barticle}[mr]
\bauthor{\bsnm{Dempster},~\bfnm{A.~P.}\binits{A.~P.}},
  \bauthor{\bsnm{Laird},~\bfnm{N.~M.}\binits{N.~M.}} \AND
  \bauthor{\bsnm{Rubin},~\bfnm{D.~B.}\binits{D.~B.}}
(\byear{1977}).
\btitle{Maximum likelihood from incomplete data via the {EM} algorithm (with discussion)}.
\bjournal{J.~Roy. Statist. Soc. Ser. B}
\bvolume{39}
\bpages{1--38}.
\bid{issn={0035-9246}, mr={0501537}}
\end{barticle}
\endbibitem


%b17 ###
\bibitem[\protect\citeauthoryear{Duerr et al.}{2006}]{Dueetal06}
%
\begin{barticle}[auto:STB|2011-03-03|12:04:44]
\bauthor{\bsnm{Duerr},~\bfnm{R.~H.}\binits{R.~H.}},
\bauthor{\bsnm{Taylor},~\bfnm{K.~D.}\binits{K.~D.}},
\bauthor{\bsnm{Brant},~\bfnm{S.~R.}\binits{S.~R.}},
\bauthor{\bsnm{Rioux},~\bfnm{J.~D.}\binits{J.~D.}},
\bauthor{\bsnm{Silverberg},~\bfnm{M.~S.}\binits{M.~S.}},
\bauthor{\bsnm{Daly},~\bfnm{M.~J.}\binits{M.~J.}},
\bauthor{\bsnm{Steinhart},~\bfnm{A.~H.}\binits{A.~H.}},
\bauthor{\bsnm{Abraham},~\bfnm{C.}\binits{C.}},
\bauthor{\bsnm{Regueiro},~\bfnm{M.}\binits{M.}} \AND
\bauthor{\bsnm{Griffiths},~\bfnm{A.}\binits{A.}}
\textsc{et al.}
(\byear{2006}).
\btitle{A genomewide association study identifies IL23R as an inflammatory
bowel disease gene}.
\bjournal{Science}
\bvolume{314}
\bpages{1461--1463}.
\end{barticle}
%
\endbibitem

%%b18 ###
%%
%(\byear{2002}).
%%

%b19 ###
\bibitem[\protect\citeauthoryear{Edenberg et al.}{2005}]{Edeetal05}
%
\begin{barticle}[auto:STB|2011-03-03|12:04:44]
\bauthor{\bsnm{Edenberg},~\bfnm{H.~J.}\binits{H.~J.}},
\bauthor{\bsnm{Bierut},~\bfnm{L.~J.}\binits{L.~J.}},
\bauthor{\bsnm{Boyce},~\bfnm{P.}\binits{P.}},
\bauthor{\bsnm{Cao},~\bfnm{M.}\binits{M.}},
\bauthor{\bsnm{Cawley},~\bfnm{S.}\binits{S.}},
\bauthor{\bsnm{Chiles},~\bfnm{R.}\binits{R.}} \AND
\bauthor{\bsnm{Doheny},~\bfnm{K.~F.}\binits{K.~F.}}
(\byear{2005}).
\btitle{Description of the data from the Collaborative Study on the
Genetics of
Alcoholism (COGA) and single-nucleotide polymorphism genotyping for Genetic
Analysis Workshop 14}.
\bjournal{BMC Genetics}
\bvolume{6}
\bpages{S2}.
\end{barticle}
%
\endbibitem

%b20 ###
\bibitem[\protect\citeauthoryear{Elston and Steward}{1971}]{ElsSte71}
%
\begin{barticle}[auto:STB|2011-03-03|12:04:44]
\bauthor{\bsnm{Elston},~\bfnm{R.~C.}\binits{R.~C.}} \AND
\bauthor{\bsnm{Steward},~\bfnm{J.}\binits{J.}}
(\byear{1971}).
\btitle{A general model for the analysis of pedigree data}.
\bjournal{Hum. Hered.}
\bvolume{21}
\bpages{523--542}.
\end{barticle}
%
\endbibitem

%%b21 ###
%%
%(\byear{1972}).
%%

%b22 ###
\bibitem[\protect\citeauthoryear{Feng, Leckman and Zhang}{2004}]{FenLecZha04}
%
\begin{barticle}[auto:STB|2011-03-03|12:04:44]
\bauthor{\bsnm{Feng},~\bfnm{R.}\binits{R.}},
\bauthor{\bsnm{Leckman},~\bfnm{J.}\binits{J.}} \AND
\bauthor{\bsnm{Zhang},~\bfnm{H.~P.}\binits{H.~P.}}
(\byear{2004}).
\btitle{Linkage analysis of ordinal traits for pedigree data}.
\bjournal{Proc. Natl. Acad. Sci. USA}
\bvolume{101}
\bpages{16739--16744}.
\end{barticle}
%
\endbibitem

%b23 ###
\bibitem[\protect\citeauthoryear{Gastwirth}{1966}]{Gas66}
%
\begin{barticle}[mr]
\bauthor{\bsnm{Gastwirth},~\bfnm{Joseph~L.}\binits{J.~L.}}
(\byear{1966}).
\btitle{On robust procedures}.
\bjournal{J. Amer. Statist. Assoc.}
\bvolume{61}
\bpages{929--948}.
\bid{issn={0162-1459}, mr={0205397}}
\end{barticle}
%
\endbibitem

%b24 ###
\bibitem[\protect\citeauthoryear{Gastwirth}{1985}]{Gas85}
%
\begin{barticle}[mr]
\bauthor{\bsnm{Gastwirth},~\bfnm{Joseph~L.}\binits{J.~L.}}
(\byear{1985}).
\btitle{The use of maximin efficiency robust tests in combining contingency
tables and survival analysis}.
\bjournal{J.~Amer. Statist. Assoc.}
\bvolume{80}
\bpages{381--384}.
\bid{issn={0162-1459}, mr={0792737}}
\end{barticle}
%
\endbibitem

%b25 ###
\bibitem[\protect\citeauthoryear{Goldgar}{1990}]{Gol90}
%
\begin{barticle}[pbm]
\bauthor{\bsnm{Goldgar},~\bfnm{D.~E.}\binits{D.~E.}}
(\byear{1990}).
\btitle{Multipoint analysis of human quantitative genetic variation}.
\bjournal{Am. J. Hum. Genet.}
\bvolume{47}
\bpages{957--967}.
\bid{issn={0002-9297}, pmcid={1683895}, pmid={2239972}}
\end{barticle}
%
\endbibitem

%b26 ###
\bibitem[\protect\citeauthoryear{Goldstein}{2009}]{Gol09}
%
\begin{barticle}[auto:STB|2011-03-03|12:04:44]
\bauthor{\bsnm{Goldstein},~\bfnm{D.~B.}\binits{D.~B.}}
(\byear{2009}).
\btitle{Common genetic variation and human traits}.
\bjournal{N. Eng. J. Med.}
\bvolume{360}
\bpages{1696--1698}.
\end{barticle}
%
\endbibitem

%b27 ###
\bibitem[\protect\citeauthoryear{Guo and Thompson}{1992}]{GuoTho92}
%
\begin{barticle}[pbm]
\bauthor{\bsnm{Guo},~\bfnm{S.~W.}\binits{S.~W.}} \AND
\bauthor{\bsnm{Thompson},~\bfnm{E.~A.}\binits{E.~A.}}
(\byear{1992}).
\btitle{A Monte Carlo method for combined segregation and linkage analysis}.
\bjournal{Am. J. Hum. Genet.}
\bvolume{51}
\bpages{1111--1126}.
\bid{issn={0002-9297}, pmcid={1682838}, pmid={1415253}}
\end{barticle}
%
\endbibitem

%b28 ###
\bibitem[\protect\citeauthoryear{Heath et al.}{2002}]{Heaetal02}
%
\begin{barticle}[pbm]
\bauthor{\bsnm{Heath},~\bfnm{Andrew~C.}\binits{A.~C.}},
\bauthor{\bsnm{Todorov},~\bfnm{Alexandre~A.}\binits{A.~A.}},
\bauthor{\bsnm{Nelson},~\bfnm{Elliot~C.}\binits{E.~C.}},
\bauthor{\bsnm{Madden},~\bfnm{Pamela A~F}\binits{P.~A.~F.}},
\bauthor{\bsnm{Bucholz},~\bfnm{Kathleen~K.}\binits{K.~K.}} \AND
\bauthor{\bsnm{Martin},~\bfnm{Nicholas~G.}\binits{N.~G.}}
(\byear{2002}).
\btitle{Gene--environment interaction effects on behavioral variation
and risk
of complex disorders: The example of alcoholism and other psychiatric
disorders}.
\bjournal{Twin Research}
\bvolume{5}
\bpages{30--37}.
\bid{issn={1369-0523}, pmid={11893279}}
\end{barticle}
%
\endbibitem

%b29 ###
\bibitem[\protect\citeauthoryear{Heston}{1966}]{Hes66}
%
\begin{barticle}[auto:STB|2011-03-03|12:04:44]
\bauthor{\bsnm{Heston},~\bfnm{L.~L.}\binits{L.~L.}}
(\byear{1966}).
\btitle{Psychiatric disorders in foster home reared children of schizophrenic
mothers}.
\bjournal{Bristish J. Psychiatry}
\bvolume{112}
\bpages{819--825}.
\end{barticle}
%
\endbibitem

%b30 ###
\bibitem[\protect\citeauthoryear{Hindorff et al.}{2009}]{Hinetal09}
%
\begin{barticle}[auto:STB|2011-03-03|12:04:44]
\bauthor{\bsnm{Hindorff},~\bfnm{L.~A.}\binits{L.~A.}},
\bauthor{\bsnm{Sethupathy},~\bfnm{P.}\binits{P.}},
\bauthor{\bsnm{Junkinsa},~\bfnm{H.~A.}\binits{H.~A.}},
\bauthor{\bsnm{Ramosa},~\bfnm{E.~M.}\binits{E.~M.}},
\bauthor{\bsnm{Mehtac},~\bfnm{J.~P.}\binits{J.~P.}},
\bauthor{\bsnm{Collinsb},~\bfnm{F.~S.}\binits{F.~S.}} \AND
\bauthor{\bsnm{Manolioa},~\bfnm{T.~A.}\binits{T.~A.}}
(\byear{2009}).
\btitle{Potential etiologic and functional implications of genome-wide
association loci for human diseases and traits}.
\bjournal{Proc. Natl. Acad. Sci. USA}
\bvolume{106}
\bpages{9362--9367}.
\end{barticle}
%
\endbibitem

%b31 ###
\bibitem[\protect\citeauthoryear{Hollander and Wolfe}{1999}]{HolWol99}
%
\begin{bbook}[mr]
\bauthor{\bsnm{Hollander},~\bfnm{Myles}\binits{M.}} \AND
\bauthor{\bsnm{Wolfe},~\bfnm{Douglas~A.}\binits{D.~A.}}
(\byear{1999}).
\btitle{Nonparametric Statistical Methods},
\bedition{2nd} ed.
\bpublisher{Wiley}, \baddress{New York}.
\bid{mr={1666064}}
\end{bbook}
%
\endbibitem

\bibitem[\protect\citeauthoryear{Hopper}{1989}]{Hop89}
\begin{barticle}[auto:STB|2011-03-03|12:04:44]
\bauthor{\bsnm{Hopper},~\bfnm{J.~L.}\binits{J.~L.}}
(\byear{1989}).
\btitle{Modelling sibship environment in the regressive logistic model for
  familial disease}.
\bjournal{Genet. Epidemiol.}
\bvolume{6}
\bpages{235--240}.
\end{barticle}
\endbibitem


%b32 ###
\bibitem[\protect\citeauthoryear{Horvath and Laird}{1998}]{HorLai98}
%
\begin{barticle}[pbm]
\bauthor{\bsnm{Horvath},~\bfnm{S.}\binits{S.}} \AND
\bauthor{\bsnm{Laird},~\bfnm{N.~M.}\binits{N.~M.}}
(\byear{1998}).
\btitle{A discordant-sibship test for disequilibrium and linkage: No
need for
parental data}.
\bjournal{Am. J. Hum. Genet.}
\bvolume{63}
\bpages{1886--1897}.
\bid{doi={10.1086/302137}, issn={0002-9297}, pii={S0002-9297(07)61632-2},
pmcid={1377659}, pmid={9837840}}
\end{barticle}
%
\endbibitem

%b33 ###
\bibitem[\protect\citeauthoryear{Kao et al.}{2008}]{autokey33}
%
\begin{bmisc}[auto:STB|2011-03-03|12:04:44]
\bauthor{\bsnm{Kao},~\bfnm{W.~H.}\binits{W.~H.}}
\textsc{et al.}
(\byear{2009}).
\bhowpublished{MYH9 is associated with nondiabetic
end-stage renal disease in African-Americans. \textit{Nat. Genet.}}
\bvolume{40}
\bpages{1185--1192}.
\end{bmisc}
%
\endbibitem

%b34 ###
\bibitem[\protect\citeauthoryear{Kety, Rosenthal and
Wender}{1978}]{KetRosWen78}
%
\begin{bincollection}[auto:STB|2011-03-03|12:04:44]
\bauthor{\bsnm{Kety},~\bfnm{S.~S.}\binits{S.~S.}},
\bauthor{\bsnm{Rosenthal},~\bfnm{D.}\binits{D.}} \AND
\bauthor{\bsnm{Wender},~\bfnm{P.}\binits{P.}}
(\byear{1978}).
\btitle{Genetic relationships within the schizophrenia spectrum:
Evidence from
adoption studies}.
In \bbooktitle{Critical Issues in Psychiatric Diagnosis}
(\beditor{\bfnm{R.~L.}\binits{R.~L.}~\bsnm{Spitzer}} \AND
\beditor{\bfnm{D.~F.}\binits{D.~F.}~\bsnm{Klein}}, eds.)
\bpages{213--223}.
\bpublisher{Raven Press}, \baddress{New York}.
\end{bincollection}
%
\endbibitem

%b35 ###
\bibitem[\protect\citeauthoryear{Klein et al.}{2005}]{Kleetal05}
%
\begin{barticle}[pbm]
\bauthor{\bsnm{Klein},~\bfnm{Robert~J.}\binits{R.~J.}},
\bauthor{\bsnm{Zeiss},~\bfnm{Caroline}\binits{C.}},
\bauthor{\bsnm{Chew},~\bfnm{Emily~Y.}\binits{E.~Y.}},
\bauthor{\bsnm{Tsai},~\bfnm{Jen-Yue}\mbox{\binits{J.-Y.}}},
\bauthor{\bsnm{Sackler},~\bfnm{Richard~S.}\binits{R.~S.}},
\bauthor{\bsnm{Haynes},~\bfnm{Chad}\binits{C.}},
\bauthor{\bsnm{Henning},~\bfnm{Alice~K.}\binits{A.~K.}},
\bauthor{\bsnm{SanGiovanni},~\bfnm{John~Paul}\binits{J.~P.}},
\bauthor{\bsnm{Mane},~\bfnm{Shrikant~M.}\binits{S.~M.}},
\bauthor{\bsnm{Mayne},~\bfnm{Susan~T.}\binits{S.~T.}},
\bauthor{\bsnm{Brac\-ken},~\bfnm{Michael~B.}\binits{M.~B.}},
\bauthor{\bsnm{Ferris},~\bfnm{Frederick~L.}\binits{F.~L.}},
\bauthor{\bsnm{Ott},~\bfnm{Jurg}\binits{J.}},
\bauthor{\bsnm{Barnstable},~\bfnm{Colin}\binits{C.}} \AND
\bauthor{\bsnm{Hoh},~\bfnm{Josephine}\binits{J.}}
(\byear{2005}).
\btitle{Complement factor H polymorphism in age-related macular degeneration}.
\bjournal{Science}
\bvolume{308}
\bpages{385--389}.
\bid{doi={10.1126/science.1109557}, issn={1095-9203}, mid={NIHMS10068},
pii={1109557}, pmcid={1512523}, pmid={15761122}}
\end{barticle}
%
\endbibitem

%b36 ###
\bibitem[\protect\citeauthoryear{Knapp}{1999}]{Kna99}
%
\begin{barticle}[pbm]
\bauthor{\bsnm{Knapp},~\bfnm{M.}\binits{M.}}
(\byear{1999}).
\btitle{Using exact $P$ values to~compare the power between the
reconstruction-combined transmission/\break disequilibrium test and the sib
transmission/disequilibrium test}.
\bjournal{Am. J. Hum. Genet.}
\bvolume{65}
\bpages{1208--1210}.
\bid{doi={10.1086/302591}, issn={0002-9297}, pii={S0002-9297(07)62630-5},
pmcid={1288260}, pmid={10486344}}
\end{barticle}
%
\endbibitem

%b37 ###
\bibitem[\protect\citeauthoryear{Kopp et al.}{2008}]{Kopetal08}
%
\begin{barticle}[pbm]
\bauthor{\bsnm{Kopp},~\bfnm{Jeffrey~B.}\binits{J.~B.}},
\bauthor{\bsnm{Smith},~\bfnm{Michael~W.}\binits{M.~W.}},
\bauthor{\bsnm{Nelson},~\bfnm{George~W.}\binits{G.~W.}},
\bauthor{\bsnm{Johnson},~\bfnm{Randall~C.}\binits{R.~C.}},
\bauthor{\bsnm{Freedman},~\bfnm{Barry~I.}\binits{B.~I.}},
\bauthor{\bsnm{Bowden},~\bfnm{Donald~W.}\binits{D.~W.}},
\bauthor{\bsnm{Olek\-syk},~\bfnm{Taras}\binits{T.}},
\bauthor{\bsnm{McKenzie},~\bfnm{Louise~M.}\binits{L.~M.}},
\bauthor{\bsnm{Kajiyama},~\bfnm{Hiroshi}\binits{H.}},
\bauthor{\bsnm{Ahuja},~\bfnm{Tejinder~S.}\binits{T.~S.}},
\bauthor{\bsnm{Berns},~\bfnm{Jeffrey~S.}\binits{J.~S.}},
\bauthor{\bsnm{Briggs},~\bfnm{William}\binits{W.}},
\bauthor{\bsnm{Cho},~\bfnm{Monique~E.}\binits{M.~E.}},
\bauthor{\bsnm{Dart},~\bfnm{Richard~A.}\binits{R.~A.}},
\bauthor{\bsnm{Kimmel},~\bfnm{Paul~L.}\binits{P.~L.}},
\bauthor{\bsnm{Korbet},~\bfnm{Stephen~M.}\binits{S.~M.}},
\bauthor{\bsnm{Michel},~\bfnm{Donna~M.}\binits{D.~M.}},
\bauthor{\bsnm{Mok\-rzycki},~\bfnm{Michele~H.}\binits{M.~H.}},
\bauthor{\bsnm{Schelling},~\bfnm{Jeffrey~R.}\binits{J.~R.}},
\bauthor{\bsnm{Simon},~\bfnm{Eric}\binits{E.}},
\bauthor{\bsnm{Trachtman},~\bfnm{Howard}\binits{H.}},
\bauthor{\bsnm{Vlahov},~\bfnm{David}\binits{D.}} \AND
\bauthor{\bsnm{Winkler},~\bfnm{Cheryl~A.}\binits{C.~A.}}
(\byear{2008}).
\btitle{MYH9 is a major-effect risk gene for focal segmental
glomerulosclerosis}.
\bjournal{Nat. Genet.}
\bvolume{40}
\bpages{1175--1184}.
\bid{doi={10.1038/ng.226}, issn={1546-1718}, mid={NIHMS167236}, pii={ng.226},
pmcid={2827354}, pmid={18794856}}
\end{barticle}
%
\endbibitem

%b38 ###
\bibitem[\protect\citeauthoryear{Kraepelin}{1899}]{Kra99}
%
\begin{bbook}[auto:STB|2011-03-03|12:04:44]
\bauthor{\bsnm{Kraepelin},~\bfnm{E.}\binits{E.}}
(\byear{1899}).
\btitle{Psychiatrie: Ein Lehrbuch fur Studirende und Aerzte}.
\bpublisher{Barth, Leipzig}.
\end{bbook}
%
\endbibitem

%b39 ###
\bibitem[\protect\citeauthoryear{Kruglyak et al.}{1996}]{Kruetal96}
%
\begin{barticle}[pbm]
\bauthor{\bsnm{Kruglyak},~\bfnm{L.}\binits{L.}},
\bauthor{\bsnm{Daly},~\bfnm{M.~J.}\binits{M.~J.}},
\bauthor{\bsnm{Reeve-Daly},~\bfnm{M.~P.}\binits{M.~P.}} \AND
\bauthor{\bsnm{Lander},~\bfnm{E.~S.}\binits{E.~S.}}
(\byear{1996}).
\btitle{Parametric and nonparametric linkage analysis: A unified multipoint
approach}.
\bjournal{Am. J. Hum. Genet.}
\bvolume{58}
\bpages{1347--1363}.
\bid{issn={0002-9297}, pmcid={1915045}, pmid={8651312}}
\end{barticle}
%
\endbibitem

%%b40 ###
%%
%(\byear{2000}).
%%

%%b41 ###
%%
%(\byear{2002}).
%proportional odds model}.
%%

%b42 ###
\bibitem[\protect\citeauthoryear{Lander and Green}{1987}]{LanGre87}
%
\begin{barticle}[pbm]
\bauthor{\bsnm{Lander},~\bfnm{E.~S.}\binits{E.~S.}} \AND
\bauthor{\bsnm{Green},~\bfnm{P.}\binits{P.}}
(\byear{1987}).
\btitle{Construction of multilocus genetic linkage maps in humans}.
\bjournal{Proc. Natl. Acad. Sci. USA}
\bvolume{84}
\bpages{2363--2367}.
\bid{issn={0027-8424}, pmcid={304651}, pmid={3470801}}
\end{barticle}
%
\endbibitem

%%b43 ###
%%
%(\byear{2002}).
%association tests: Quantitative traits}.
%pmcid={378574}, pmid={12454799}}
%%

%%b44 ###
%%
%(\byear{2002}).
%tests: Dichotomous traits}.
%pmcid={379194}, pmid={12181775}}
%%

%b45 ###
\bibitem[\protect\citeauthoryear{Lange et al.}{2003}]{Lanetal03}
%
\begin{barticle}[auto:STB|2011-03-03|12:04:44]
\bauthor{\bsnm{Lange},~\bfnm{C.}\binits{C.}},
\bauthor{\bsnm{Silverman},~\bfnm{E.~K.}\binits{E.~K.}},
\bauthor{\bsnm{Xu},~\bfnm{X.}\binits{X.}},
\bauthor{\bsnm{Weiss},~\bfnm{S.~T.}\binits{S.~T.}} \AND
\bauthor{\bsnm{Laird},~\bfnm{N.~M.}\binits{N.~M.}}
(\byear{2003}).
\btitle{A multivariate family-based association test using generalized
estimating equations: FBAT-GEE}.
\bjournal{Biostatistics}
\bvolume{4}
\bpages{195--306}.
\end{barticle}
%
\endbibitem

\bibitem[\protect\citeauthoryear{Li and Thompson}{1997}]{LiTho97}
\begin{barticle}[auto:STB|2011-03-03|12:04:44]
\bauthor{\bsnm{Li},~\bfnm{H.~Z.}\binits{H.~Z.}} \AND
  \bauthor{\bsnm{Thompson},~\bfnm{E.}\binits{E.}}
(\byear{1997}).
\btitle{Semiparametric estimation of major gene and family-specific random
  effects for age of onset}.
\bjournal{Biometrics}
\bvolume{53}
\bpages{282--293}.
\end{barticle}
\endbibitem


%b46 ###
\bibitem[\protect\citeauthoryear{Li and Burmeister}{2009}]{LiBur09}
%
\begin{barticle}[pbm]
\bauthor{\bsnm{Li},~\bfnm{Ming~D.}\binits{M.~D.}} \AND
\bauthor{\bsnm{Burmeister},~\bfnm{Margit}\binits{M.}}
(\byear{2009}).
\btitle{New insights into the genetics of addiction}.
\bjournal{Nat. Rev. Genet.}
\bvolume{10}
\bpages{225--231}.
\bid{doi={10.1038/nrg2536}, issn={1471-0064}, mid={NIHMS200025}, pii={nrg2536},
pmcid={2879628}, pmid={19238175}}
\end{barticle}
%
\endbibitem

\bibitem[\protect\citeauthoryear{Liang and Rathouz}{1999}]{LiaRat99}
\begin{barticle}[mr]
\bauthor{\bsnm{Liang},~\bfnm{Kung-Yee}\binits{K.-Y.}} \AND
  \bauthor{\bsnm{Rathouz},~\bfnm{Paul~J.}\binits{P.~J.}}
(\byear{1999}).
\btitle{Hypothesis testing under mixture models: Application to genetic linkage
  analysis}.
\bjournal{Biometrics}
\bvolume{55}
\bpages{65--74}.
\bid{doi={10.1111/j.0006-341X.1999.00065.x}, issn={0006-341X}, mr={1705673}}
\end{barticle}
\endbibitem

%b47 ###
\bibitem[\protect\citeauthoryear{Liu, Tang and Zhang}{2009}]{LiuTanZha09}
%
\begin{barticle}[mr]
\bauthor{\bsnm{Liu},~\bfnm{Huan}\binits{H.}},
\bauthor{\bsnm{Tang},~\bfnm{Yongqiang}\binits{Y.}} \AND
\bauthor{\bsnm{Zhang},~\bfnm{Hao~Helen}\binits{H.~H.}}
(\byear{2009}).
\btitle{A new chi-square approximation to the distribution of non-negative
definite quadratic forms in non-central normal variables}.
\bjournal{Comput. Statist. Data Anal.}
\bvolume{53}
\bpages{853--856}.
\bid{doi={10.1016/j.csda.2008.11.025}, issn={0167-9473}, mr={2657050}}
\end{barticle}
%
\endbibitem

%b48 ###
\bibitem[\protect\citeauthoryear{Liu, Tritchler and Bull}{2002}]{LiuTriBul02}
%
\begin{barticle}[auto:STB|2011-03-03|12:04:44]
\bauthor{\bsnm{Liu},~\bfnm{Y.}\binits{Y.}},
\bauthor{\bsnm{Tritchler},~\bfnm{D.}\binits{D.}} \AND
\bauthor{\bsnm{Bull},~\bfnm{S.~B.}\binits{S.~B.}}
(\byear{2002}).
\btitle{A unified framework for transmission-disequilibrium test
analysis of
discrete and continuous traits}.
\bjournal{Genet. Epidemiol.}
\bvolume{22}
\bpages{26--40}.
\end{barticle}
%
\endbibitem

%b49 ###
\bibitem[\protect\citeauthoryear{Lunetta et al.}{2000}]{Lunetal00}
%
\begin{barticle}[auto:STB|2011-03-03|12:04:44]
\bauthor{\bsnm{Lunetta},~\bfnm{K.~L.}\binits{K.~L.}},
\bauthor{\bsnm{Farone},~\bfnm{S.~V.}\binits{S.~V.}},
\bauthor{\bsnm{Biederman},~\bfnm{J.}\binits{J.}} \AND
\bauthor{\bsnm{Laird},~\bfnm{N.~M.}\binits{N.~M.}}
(\byear{2000}).
\btitle{Family based tests of association and linkage that used unaffected
sibs, covariates, and interactions}.
\bjournal{Am. J. Hum. Genet.}
\bvolume{66}
\bpages{605--614}.
\end{barticle}
%
\endbibitem

%b50 ###
\bibitem[\protect\citeauthoryear{Martin et al.}{2000}]{Maretal00}
%
\begin{barticle}[pbm]
\bauthor{\bsnm{Martin},~\bfnm{E.~R.}\binits{E.~R.}},
\bauthor{\bsnm{Monks},~\bfnm{S.~A.}\binits{S.~A.}},
\bauthor{\bsnm{Warren},~\bfnm{L.~L.}\binits{L.~L.}} \AND
\bauthor{\bsnm{Kaplan},~\bfnm{N.~L.}\binits{N.~L.}}
(\byear{2000}).
\btitle{A test for linkage and association in general pedigrees: The pedigree
disequilibrium test}.
\bjournal{Am. J. Hum. Genet.}
\bvolume{67}
\bpages{146--154}.
\bid{doi={10.1086/302957}, issn={0002-9297}, pii={S0002-9297(07)62440-9},
pmcid={1287073}, pmid={10825280}}
\end{barticle}
%
\endbibitem

%b51 ###
\bibitem[\protect\citeauthoryear{Merikangas et al.}{1998}]{Meretal98}
%
\begin{barticle}[pbm]
\bauthor{\bsnm{Merikangas},~\bfnm{K.~R.}\binits{K.~R.}},
\bauthor{\bsnm{Stolar},~\bfnm{M.}\binits{M.}},
\bauthor{\bsnm{Stevens},~\bfnm{D.~E.}\binits{D.~E.}},
\bauthor{\bsnm{Goulet},~\bfnm{J.}\binits{J.}},
\bauthor{\bsnm{Preisig},~\bfnm{M.~A.}\binits{M.~A.}},
\bauthor{\bsnm{Fenton},~\bfnm{B.}\binits{B.}},
\bauthor{\bsnm{Zhang},~\bfnm{H.}\binits{H.}},
\bauthor{\bsnm{O'Malley},~\bfnm{S.~S.}\binits{S.~S.}} \AND
\bauthor{\bsnm{Rounsaville},~\bfnm{B.~J.}\binits{B.~J.}}
(\byear{1998}).
\btitle{Familial transmission of substance use disorders}.
\bjournal{Arch. Gen. Psychiatry}
\bvolume{55}
\bpages{973--979}.
\bid{issn={0003-990X}, pmid={9819065}}
\end{barticle}
%
\endbibitem

%b52 ###
\bibitem[\protect\citeauthoryear{Morton}{1955}]{MOR55}
%
\begin{barticle}[pbm]
\bauthor{\bsnm{Morton},~\bfnm{N.~E.}\binits{N.~E.}}
(\byear{1955}).
\btitle{Sequential tests for the detection of linkage}.
\bjournal{Am. J. Hum. Genet.}
\bvolume{7}
\bpages{277--318}.
\bid{issn={0002-9297}, pmcid={1716611}, pmid={13258560}}
\end{barticle}
%
\endbibitem

%b53 ###
\bibitem[\protect\citeauthoryear{Mulligan et al.}{2006}]{Muletal06}
%
\begin{barticle}[pbm]
\bauthor{\bsnm{Mulligan},~\bfnm{Megan~K.}\binits{M.~K.}},
\bauthor{\bsnm{Ponomarev},~\bfnm{Igor}\binits{I.}},
\bauthor{\bsnm{Hitzemann},~\bfnm{Robert~J.}\binits{R.~J.}},
\bauthor{\bsnm{Belknap},~\bfnm{John~K.}\binits{J.~K.}},
\bauthor{\bsnm{Tabakoff},~\bfnm{Boris}\binits{B.}},
\bauthor{\bsnm{Harris},~\bfnm{R.~Adron}\binits{R.~A.}},
\bauthor{\bsnm{Crabbe},~\bfnm{John~C.}\binits{J.~C.}},
\bauthor{\bsnm{Blednov},~\bfnm{Yuri~A.}\binits{Y.~A.}},
\bauthor{\bsnm{Grahame},~\bfnm{Nicholas~J.}\binits{N.~J.}},
\bauthor{\bsnm{Phillips},~\bfnm{Tamara~J.}\binits{T.~J.}},
\bauthor{\bsnm{Finn},~\bfnm{Deborah~A.}\binits{D.~A.}},
\bauthor{\bsnm{Hoffman},~\bfnm{Paula~L.}\binits{P.~L.}},
\bauthor{\bsnm{Iyer},~\bfnm{Vishwanath~R.}\binits{V.~R.}},
\bauthor{\bsnm{Koob},~\bfnm{George~F.}\binits{G.~F.}} \AND
\bauthor{\bsnm{Bergeson},~\bfnm{Susan~E.}\binits{S.~E.}}
(\byear{2006}).
\btitle{Toward understanding the genetics of alcohol drinking through
transcriptome meta-analysis}.
\bjournal{Proc. Natl. Acad. Sci. USA}
\bvolume{103}
\bpages{6368--6373}.
\bid{doi={10.1073/pnas.0510188103}, issn={0027-8424}, pii={0510188103},
pmcid={1458884}, pmid={16618939}}
\end{barticle}
%
\endbibitem

%b54 ###
\bibitem[\protect\citeauthoryear{Ott}{1974}]{Ott74}
%
\begin{barticle}[pbm]
\bauthor{\bsnm{Ott},~\bfnm{J.}\binits{J.}}
(\byear{1974}).
\btitle{Estimation of the recombination fraction in human pedigrees: Efficient
computation of the likelihood for human linkage studies}.
\bjournal{Am. J. Hum. Genet.}
\bvolume{26}
\bpages{588--597}.
\bid{issn={0002-9297}, pmcid={1762722}, pmid={4422075}}
\end{barticle}
%
\endbibitem

%b55 ###
\bibitem[\protect\citeauthoryear{Ott}{1999}]{Ott99}
%
\begin{bbook}[auto:STB|2011-03-03|12:04:44]
\bauthor{\bsnm{Ott},~\bfnm{J.}\binits{J.}}
(\byear{1999}).
\btitle{Analysis of Human Genetic Linkage}, \bedition{3rd} ed.
\bpublisher{Johns Hopkins Univ. Press}, \baddress{Baltimore, MD}.
\end{bbook}
%
\endbibitem

%b56 ###
\bibitem[\protect\citeauthoryear{Pauls and Leckman}{1986}]{PauLec86}
%
\begin{barticle}[auto:STB|2011-03-03|12:04:44]
\bauthor{\bsnm{Pauls},~\bfnm{D.~L.}\binits{D.~L.}} \AND
\bauthor{\bsnm{Leckman},~\bfnm{J.~F.}\binits{J.~F.}}
(\byear{1986}).
\btitle{The inheritance of Gilles de la Tourette's syndrome and associated
behaviors. Evidence for autosomal dominant transmission}.
\bjournal{New Eng. J. Med.}
\bvolume{315}
\bpages{993--997}.
\end{barticle}
%
\endbibitem

%b57 ###
\bibitem[\protect\citeauthoryear{Pearson}{1959}]{Pea59}
%
\begin{barticle}[mr]
\bauthor{\bsnm{Pearson},~\bfnm{E.~S.}\binits{E.~S.}}
(\byear{1959}).
\btitle{Note on an approximation to the distribution of non-central
{$\chi
\sp{2}$}}.
\bjournal{Biometrika}
\bvolume{46}
\bpages{364}.
\bid{issn={0006-3444}, mr={0109380}}
\end{barticle}
%
\endbibitem

%b58 ###
\bibitem[\protect\citeauthoryear{Plomin et al.}{1997}]{Ploetal97}
%
\begin{bbook}[auto:STB|2011-03-03|12:04:44]
\bauthor{\bsnm{Plomin},~\bfnm{R.}\binits{R.}},
\bauthor{\bsnm{DeFries},~\bfnm{J.~C.}\binits{J.~C.}},
\bauthor{\bsnm{McClearn},~\bfnm{G.~E.}\binits{G.~E.}} \AND
\bauthor{\bsnm{Rutter},~\bfnm{M.}\binits{M.}}
(\byear{1997}).
\btitle{Behavioral Genetics}, \bedition{3rd} ed.
\bpublisher{Freeman}, \baddress{New York}.
\end{bbook}
%
\endbibitem

%b59 ###
\bibitem[\protect\citeauthoryear{Purcell et al.}{2007}]{Puretal07}
%
\begin{barticle}[auto:STB|2011-03-03|12:04:44]
\bauthor{\bsnm{Purcell},~\bfnm{S.}\binits{S.}},
\bauthor{\bsnm{Neale},~\bfnm{B.}\binits{B.}},
\bauthor{\bsnm{Todd-Brown},~\bfnm{K.}\binits{K.}},
\bauthor{\bsnm{Thomas},~\bfnm{L.}\binits{L.}},
\bauthor{\bsnm{Ferreira},~\bfnm{M.~A.~R.}\binits{M.~A.~R.}},
\bauthor{\bsnm{Bender},~\bfnm{D.}\binits{D.}},
\bauthor{\bsnm{Maller},~\bfnm{J.}\binits{J.}},
\bauthor{\bsnm{Sklar},~\bfnm{P.}\binits{P.}}, \bauthor{\bparticle{de}
\bsnm{Bakker},~\bfnm{P.~I.~W.}\binits{P.~I.~W.}},
\bauthor{\bsnm{Daly},~\bfnm{M.~J.}\binits{M.~J.}} \AND
\bauthor{\bsnm{Sham},~\bfnm{P.~C.}\binits{P.~C.}}
(\byear{2007}).
\btitle{PLINK: A toolset for whole-genome association and population-based
linkage analysis}.
\bjournal{Am. J. Hum. Genet.}
\bvolume{81}
\bpages{559--575}.
\end{barticle}
%
\endbibitem

%b60 ###
\bibitem[\protect\citeauthoryear{Rabinowitz}{1997}]{Rab}
%
\begin{barticle}[pbm]
\bauthor{\bsnm{Rabinowitz},~\bfnm{D.}\binits{D.}}
(\byear{1997}).
\btitle{A transmission disequilibrium test for quantitative trait loci}.
\bjournal{Hum. Hered.}
\bvolume{47}
\bpages{342--350}.
\bid{issn={0001-5652}, pmid={9391826}}
\end{barticle}
%
\endbibitem

%b61 ###
\bibitem[\protect\citeauthoryear{Rabinowitz and Laird}{2000}]{RabLai00}
%
\begin{barticle}[auto:STB|2011-03-03|12:04:44]
\bauthor{\bsnm{Rabinowitz},~\bfnm{D.}\binits{D.}} \AND
\bauthor{\bsnm{Laird},~\bfnm{N.~M.}\binits{N.~M.}}
(\byear{2000}).
\btitle{A unified approach to adjusting association tests for population
admixture with arbitrary pedigree structure and arbitrary missing marker
information}.
\bjournal{Hum. Hered.}
\bvolume{50}
\bpages{211--223}.
\end{barticle}
%
\endbibitem

%b62 ###
\bibitem[\protect\citeauthoryear{Rao and Xu}{1998}]{RaoXu98}
%
\begin{barticle}[pbm]
\bauthor{\bsnm{Rao},~\bfnm{S.}\binits{S.}} \AND
\bauthor{\bsnm{Xu},~\bfnm{S.}\binits{S.}}
(\byear{1998}).
\btitle{Mapping quantitative trait loci for ordered categorical traits in
four-way crosses}.
\bjournal{Heredity}
\bvolume{81}
\bpages{214--224}.
\bid{issn={0018-067X}, pmid={9750263}}
\end{barticle}
%
\endbibitem

%%b63 ###
%%
%(\byear{1998}).
%dependence}.
%pii={10.1002/(SICI)1096-8628(19980508)81:3<207::AID-AJMG1>3.0.CO;2-T},
%pmid={9603606}}
%%

%b64 ###
\bibitem[\protect\citeauthoryear{Risch and Zhang}{1995}]{RisZha95}
%
\begin{barticle}[auto:STB|2011-03-03|12:04:44]
\bauthor{\bsnm{Risch},~\bfnm{N.~R.}\binits{N.~R.}} \AND
\bauthor{\bsnm{Zhang},~\bfnm{H.~P.}\binits{H.~P.}}
(\byear{1995}).
\btitle{Extreme discordant sib pairs for mapping quantitative trait
loci in
humans}.
\bjournal{Science}
\bvolume{268}
\bpages{1584--1589}.
\end{barticle}
%
\endbibitem

%b65 ###
\bibitem[\protect\citeauthoryear{Rosenthal}{1972}]{Ros72}
%
\begin{bmisc}[auto:STB|2011-03-03|12:04:44]
\bauthor{\bsnm{Rosenthal},~\bfnm{D.}\binits{D.}}
(\byear{1972}).
\bhowpublished{Three adoption studies of heredity in the schizophrenic disorders.
\textit{Internat. J. Mental Health}}
\bvolume{1}
\bpages{63--75}.
\end{bmisc}
%
\endbibitem

%b66 ###
\bibitem[\protect\citeauthoryear{Scharf et al.}{2008}]{Schetal08}
%
\begin{bmisc}[pbm]
\bauthor{\bsnm{Scharf},~\bfnm{J.~M.}\binits{J.~M.}},
\bauthor{\bsnm{Moorjani},~\bfnm{P.}\binits{P.}},
\bauthor{\bsnm{Fagerness},~\bfnm{J.}\binits{J.}},
\bauthor{\bsnm{Plat\-ko},~\bfnm{J.~V.}\binits{J.~V.}},
\bauthor{\bsnm{Illmann},~\bfnm{C.}\binits{C.}},
\bauthor{\bsnm{Galloway},~\bfnm{B.}\binits{B.}},
\bauthor{\bsnm{Jenike},~\bfnm{E.}\binits{E.}},
\bauthor{\bsnm{Stewart},~\bfnm{S.~E.}\binits{S.~E.}},
\bauthor{\bsnm{Pauls},~\bfnm{D.~L.}\binits{D.~L.}} \AND
\bauthor{\bparticle{The Tourette Syndrome International Consortium for Genetics}}
(\byear{2008}).
\bhowpublished{Lack of association between SLITRK1var321 and Tourette
syndrome in a
large family-based sample.
\textit{Neurology}}
\bvolume{70}
\bpages{1495--1496}.
\bid{doi={10.1212/01.wnl.0000296833.25484.bb}, issn={1526-632X},
pii={70/16_Part_2/1495}, pmid={18413575}}
\end{bmisc}
%
\endbibitem

%b67 ###
\bibitem[\protect\citeauthoryear{Schork}{1993}]{Sch93}
%
\begin{barticle}[auto:STB|2011-03-03|12:04:44]
\bauthor{\bsnm{Schork},~\bfnm{N.~J.}\binits{N.~J.}}
(\byear{1993}).
\btitle{Extended multipoint identity-by-descent analy-sis of human quantitative
traits: Efficiency, power, and modeling con-siderations}.
\bjournal{Am. J. Hum. Genet.}
\bvolume{53}
\bpages{1306--1319}.
\end{barticle}
%
\endbibitem

\bibitem[\protect\citeauthoryear{Siegmund and McKnight}{1998}]{SieMcK98}
\begin{barticle}[auto:STB|2011-03-03|12:04:44]
\bauthor{\bsnm{Siegmund},~\bfnm{K.}\binits{K.}} \AND
  \bauthor{\bsnm{McKnight},~\bfnm{B.}\binits{B.}}
(\byear{1998}).
\btitle{Modeling hazard functions in families}.
\bjournal{Genet. Epidemiol.}
\bvolume{15}
\bpages{147--171}.
\end{barticle}
\endbibitem

%b68 ###
\bibitem[\protect\citeauthoryear{Solomon and Stephens}{1977}]{SolSte77}
%
\begin{barticle}[auto:STB|2011-03-03|12:04:44]
\bauthor{\bsnm{Solomon},~\bfnm{H.}\binits{H.}} \AND
\bauthor{\bsnm{Stephens},~\bfnm{M.~A.}\binits{M.~A.}}
(\byear{1977}).
\btitle{Distribution of a sum of weighted chi-square variables}.
\bjournal{J. Amer. Statist. Assoc.}
\bvolume{72}
\bpages{881--885}.
\end{barticle}
%
\endbibitem

%b69 ###
\bibitem[\protect\citeauthoryear{Spielman and Ewens}{1998}]{SpiEwe98}
%
\begin{barticle}[auto:STB|2011-03-03|12:04:44]
\bauthor{\bsnm{Spielman},~\bfnm{R.~S.}\binits{R.~S.}} \AND
\bauthor{\bsnm{Ewens},~\bfnm{W.~J.}\binits{W.~J.}}
(\byear{1998}).
\btitle{A sibship rest for linkage in the presence of association: The sib
transmission/disequilibrium test}.
\bjournal{Am. J. Hum. Genet.}
\bvolume{62}
\bpages{450--458}.
\end{barticle}
%
\endbibitem

%b70 ###
\bibitem[\protect\citeauthoryear{Spielman, McGinnis and
Ewens}{1993}]{SpiMcGEwe93}
%
\begin{barticle}[pbm]
\bauthor{\bsnm{Spielman},~\bfnm{R.~S.}\binits{R.~S.}},
\bauthor{\bsnm{McGinnis},~\bfnm{R.~E.}\binits{R.~E.}} \AND
\bauthor{\bsnm{Ewens},~\bfnm{W.~J.}\binits{W.~J.}}
(\byear{1993}).
\btitle{Transmission test for linkage disequilibrium: The insulin gene region
and insulin-dependent diabetes mellitus (IDDM)}.
\bjournal{Am. J. Hum. Genet.}
\bvolume{52}
\bpages{506--516}.
\bid{issn={0002-9297}, pmcid={1682161}, pmid={8447318}}
\end{barticle}
%
\endbibitem

%b71 ###
\bibitem[\protect\citeauthoryear{Steinke, Borish and
Rosenwasser}{2003}]{SteBorRos03}
%
\begin{barticle}[pbm]
\bauthor{\bsnm{Steinke},~\bfnm{John~W.}\binits{J.~W.}},
\bauthor{\bsnm{Borish},~\bfnm{Larry}\binits{L.}} \AND
\bauthor{\bsnm{Rosenwasser},~\bfnm{Lanny~J.}\binits{L.~J.}}
(\byear{2003}).
\btitle{Genetics of hypersensitivity}.
\bjournal{J. Allergy Clin. Immunol.}
\bvolume{111}
\bpages{S495--S501}.
\bid{issn={0091-6749}, pii={S0091674902912625}, pmid={12592296}}
\end{barticle}
%
\endbibitem

%b72 ###
\bibitem[\protect\citeauthoryear{Tienari}{1991}]{Tie91}
%
\begin{barticle}[pbm]
\bauthor{\bsnm{Tienari},~\bfnm{P.}\binits{P.}}
(\byear{1991}).
\btitle{Interaction between genetic vulnerability and family
environment: The
Finnish adoptive family study of schizophrenia}.
\bjournal{Acta Psychiatr. Scand.}
\bvolume{84}
\bpages{460--465}.
\bid{issn={0001-690X}, pmid={1776499}}
\end{barticle}
%
\endbibitem

%b73 ###
\bibitem[\protect\citeauthoryear{True et al.}{1999}]{Truetal99}
%
\begin{barticle}[pbm]
\bauthor{\bsnm{True},~\bfnm{W.~R.}\binits{W.~R.}},
\bauthor{\bsnm{Heath},~\bfnm{A.~C.}\binits{A.~C.}},
\bauthor{\bsnm{Scherrer},~\bfnm{J.~F.}\binits{J.~F.}},
\bauthor{\bsnm{Xian},~\bfnm{H.}\binits{H.}},
\bauthor{\bsnm{Lin},~\bfnm{N.}\binits{N.}},
\bauthor{\bsnm{Eisen},~\bfnm{S.~A.}\binits{S.~A.}},
\bauthor{\bsnm{Lyons},~\bfnm{M.~J.}\binits{M.~J.}},
\bauthor{\bsnm{Goldberg},~\bfnm{J.}\binits{J.}} \AND
\bauthor{\bsnm{Tsuang},~\bfnm{M.~T.}\binits{M.~T.}}
(\byear{1999}).
\btitle{Interrelationship of genetic and environmental influences on conduct
disorder and alcohol and marijuana dependence symptoms}.
\bjournal{Am. J. Med. Genet.}
\bvolume{88}
\bpages{391--397}.
\bid{issn={0148-7299},
pii={10.1002/(SICI)1096-8628(19990820)88:4<391::AID-AJMG17>3.0.CO;2-L},
pmid={10402507}}
\end{barticle}
%
\endbibitem

%b74 ###
\bibitem[\protect\citeauthoryear{Vergne et al.}{2003}]{Veretal03}
%
\begin{barticle}[auto:STB|2011-03-03|12:04:44]
\bauthor{\bsnm{Vergne},~\bfnm{L.}\binits{L.}},
\bauthor{\bsnm{Bourgeois},~\bfnm{A.}\binits{A.}},
\bauthor{\bsnm{Mpoudi-Ngole},~\bfnm{E.}\binits{E.}},
\bauthor{\bsnm{Mougnu\-tou},~\bfnm{R.}\binits{R.}},
\bauthor{\bsnm{Mbuagbaw},~\bfnm{J.}\binits{J.}},
\bauthor{\bsnm{Liegeois},~\bfnm{F.}\binits{F.}},
\bauthor{\bsnm{Laurent},~\bfnm{C.}\binits{C.}},
\bauthor{\bsnm{Butel},~\bfnm{C.}\binits{C.}},
\bauthor{\bsnm{Zekeng},~\bfnm{L.}\binits{L.}},
\bauthor{\bsnm{Delaporte},~\bfnm{E.}\binits{E.}} \AND
\bauthor{\bsnm{Peeters},~\bfnm{M.}\binits{M.}}
(\byear{2003}).
\btitle{Biological and genetic characteristics of HIV infections in Cameroon
reveals dual group M and O infections and a correlation between SI-inducing
phenotype of the predominant CRF02\_AG variant and disease stage}.
\bjournal{Virology}
\bvolume{310}
\bpages{254--266}.
\end{barticle}
%
\endbibitem

%b75 ###
\bibitem[\protect\citeauthoryear{Wang, Ye and Zhang}{2006}]{WanYeZha06}
%
\begin{barticle}[auto:STB|2011-03-03|12:04:44]
\bauthor{\bsnm{Wang},~\bfnm{X.~Q.}\binits{X.~Q.}},
\bauthor{\bsnm{Ye},~\bfnm{Y.~Q.}\binits{Y.~Q.}} \AND
\bauthor{\bsnm{Zhang},~\bfnm{H.~P.}\binits{H.~P.}}
(\byear{2006}).
\btitle{Family-based association tests for ordinal traits adjusting for
covariates}.
\bjournal{Genet. Epidemiol.}
\bvolume{30}
\bpages{728--736}.
\end{barticle}
%
\endbibitem

%b76 ###
\bibitem[\protect\citeauthoryear{Weinberg}{1999}]{Wei99}
%
\begin{barticle}[auto:STB|2011-03-03|12:04:44]
\bauthor{\bsnm{Weinberg},~\bfnm{C.~R.}\binits{C.~R.}}
(\byear{1999}).
\btitle{Allowing for missing parents in genetic studies of case-parental
triads}.
\bjournal{Am. J. Hum. Genet.}
\bvolume{64}
\bpages{1186--1193}.
\end{barticle}
%
\endbibitem

%b77 ###
\bibitem[\protect\citeauthoryear{Whittemore}{1996}]{Whi96}
%
\begin{barticle}[pbm]
\bauthor{\bsnm{Whittemore},~\bfnm{A.~S.}\binits{A.~S.}}
(\byear{1996}).
\btitle{Genome scanning for linkage: An overview}.
\bjournal{Am. J. Hum. Genet.}
\bvolume{59}
\bpages{704--716}.
\bid{issn={0002-9297}, pmcid={1914900}, pmid={8751872}}
\end{barticle}
%
\endbibitem

%b78 ###
\bibitem[\protect\citeauthoryear{Xu and Xu}{2006}]{XuXu06}
%
\begin{barticle}[pbm]
\bauthor{\bsnm{Xu},~\bfnm{S.}\binits{S.}} \AND
\bauthor{\bsnm{Xu},~\bfnm{C.}\binits{C.}}
(\byear{2006}).
\btitle{A multivariate model for ordinal trait analysis}.
\bjournal{Heredity}
\bvolume{97}
\bpages{409--417}.
\bid{doi={10.1038/sj.hdy.6800885}, issn={0018-067X}, pii={6800885},
pmid={16912701}}
\end{barticle}
%
\endbibitem

%b79 ###
\bibitem[\protect\citeauthoryear{Zhang, Feng and Zhu}{2003}]{ZhaFenZhu03}
%
\begin{barticle}[mr]
\bauthor{\bsnm{Zhang},~\bfnm{Heping}\binits{H.~P.}},
\bauthor{\bsnm{Feng},~\bfnm{Rui}\binits{R.}} \AND
\bauthor{\bsnm{Zhu},~\bfnm{Hongtu}\binits{H.}}
(\byear{2003}).
\btitle{A latent variable model of segregation analysis for ordinal traits}.
\bjournal{J.~Amer. Statist. Assoc.}
\bvolume{98}
\bpages{1023--1034}.
\bid{doi={10.1198/016214503000000981}, issn={0162-1459}, mr={2041490}}
\end{barticle}
%
\endbibitem

%b80 ###
\bibitem[\protect\citeauthoryear{Zhang, Liu and Wang}{2010}]{ZhaLiuWan10}
%
\begin{barticle}[mr]
\bauthor{\bsnm{Zhang},~\bfnm{Heping}\binits{H.~P.}},
\bauthor{\bsnm{Liu},~\bfnm{Ching-Ti}\binits{C.-T.}} \AND
\bauthor{\bsnm{Wang},~\bfnm{Xueqin}\binits{X.}}
(\byear{2010}).
\btitle{An association test for multiple traits based on the generalized
{K}endall's tau}.
\bjournal{J.~Amer. Statist. Assoc.}
\bvolume{105}
\bpages{473--481}.
\bid{doi={10.1198/jasa.2009.ap08387}, issn={0162-1459}, mr={2724840}}
\end{barticle}
%
\endbibitem

%b81 ###
\bibitem[\protect\citeauthoryear{Zhang and Merikangas}{2000}]{ZhaMer00}
%
\begin{barticle}[auto:STB|2011-03-03|12:04:44]
\bauthor{\bsnm{Zhang},~\bfnm{H.~P.}\binits{H.~P.}} \AND
\bauthor{\bsnm{Merikangas},~\bfnm{K.}\binits{K.}}
(\byear{2000}).
\btitle{A frailty model of segregation analysis: Understanding the familial
transmission of alcoholism}.
\bjournal{Biometrics}
\bvolume{56}
\bpages{815--823}.
\end{barticle}
%
\endbibitem

%b82 ###
\bibitem[\protect\citeauthoryear{Zhang, Wang and Ye}{2006}]{ZhaWanYe06}
%
\begin{barticle}[auto:STB|2011-03-03|12:04:44]
\bauthor{\bsnm{Zhang},~\bfnm{H.~P.}\binits{H.~P.}},
\bauthor{\bsnm{Wang},~\bfnm{X.~Q.}\binits{X.~Q.}} \AND
\bauthor{\bsnm{Ye},~\bfnm{Y.~Q.}\binits{Y.~Q.}}
(\byear{2006}).
\btitle{Detection of genes for ordinal traits in nuclear families and
a unified
approach for association studies}.
\bjournal{Genetics}
\bvolume{172}
\bpages{693--699}.
\end{barticle}
%
\endbibitem\

%b83 ###
\bibitem[\protect\citeauthoryear{Zhang et al.}{2008}]{Zhaetal08}
%
\begin{barticle}[pbm]
\bauthor{\bsnm{Zhang},~\bfnm{Meizhuo}\binits{M.}},
\bauthor{\bsnm{Feng},~\bfnm{Rui}\binits{R.}},
\bauthor{\bsnm{Chen},~\bfnm{Xiang}\binits{X.}},
\bauthor{\bsnm{Hu},~\bfnm{Buqu}\binits{B.}} \AND
\bauthor{\bsnm{Zhang},~\bfnm{Heping}\binits{H.}}
(\byear{2008}).
\btitle{LOT: A tool for linkage analysis of ordinal traits for
pedigree data}.
\bjournal{Bioinformatics}
\bvolume{24}
\bpages{1737--1739}.
\bid{doi={10.1093/bioinformatics/btn258}, issn={1367-4811}, mid={NIHMS53387},
pii={btn258}, pmcid={2566542}, pmid={18535081}}
\end{barticle}
%
\endbibitem

%b84 ###
\bibitem[\protect\citeauthoryear{Zheng et al.}{2009}]{Zheetal09}
%
\begin{barticle}[auto:STB|2011-03-03|12:04:44]
\bauthor{\bsnm{Zheng},~\bfnm{G.}\binits{G.}},
\bauthor{\bsnm{Joo},~\bfnm{J.}\binits{J.}},
\bauthor{\bsnm{Zaykin},~\bfnm{D.}\binits{D.}},
\bauthor{\bsnm{Wu},~\bfnm{C.}\binits{C.}} \AND
\bauthor{\bsnm{Geller},~\bfnm{N.}\binits{N.}}
(\byear{2009}).
\btitle{Robust tests in genome-wide scans under incomplete linkage
disequilibrium}.
\bjournal{Statist. Sci.}
\bvolume{24}
\bpages{503--516}.
\bid{mr={2779340}}
\end{barticle}
%
\endbibitem

%b85 ###
\bibitem[\protect\citeauthoryear{Zhu, Jiang and Zhang}{2010}]{ZhuJiaZha}
%
\begin{bmisc}[auto:STB|2011-03-03|12:04:44]
\bauthor{\bsnm{Zhu},~\bfnm{W.~S.}\binits{W.~S.}},
\bauthor{\bsnm{Jiang},~\bfnm{Y.}\binits{Y.}} \AND
\bauthor{\bsnm{Zhang},~\bfnm{H.~P.}\binits{H.~P.}}
(\byear{2010}).
\bhowpublished{Covariate-adjusted association tests and power
calculations based on the generalized Kendall's tau. Technical report}.
\end{bmisc}
%
\endbibitem

%b86 ###
\bibitem[\protect\citeauthoryear{Zhu and Zhang}{2009}]{ZhuZha09}
%
\begin{barticle}[mr]
\bauthor{\bsnm{Zhu},~\bfnm{Wensheng}\binits{W.}} \AND
\bauthor{\bsnm{Zhang},~\bfnm{Heping}\binits{H.}}
(\byear{2009}).
\btitle{Why do we test multiple traits in genetic association studies?}
\bjournal{J. Korean Statist. Soc.}
\bvolume{38}
\bpages{1--10}.
\bid{doi={10.1016/j.jkss.2008.10.006}, issn={1226-3192}, mr={2656857}}
\bptnote{check related}%
\end{barticle}
%
\endbibitem

%b87 ###
\bibitem[\protect\citeauthoryear{Zhu et al.}{2005}]{Zhuetal05}
%
\begin{barticle}[pbm]
\bauthor{\bsnm{Zhu},~\bfnm{Xiaofeng}\binits{X.}},
\bauthor{\bsnm{Cooper},~\bfnm{Richard}\binits{R.}},
\bauthor{\bsnm{Kan},~\bfnm{Donghui}\binits{D.}},
\bauthor{\bsnm{Cao},~\bfnm{Guichan}\binits{G.}} \AND
\bauthor{\bsnm{Wu},~\bfnm{Xiaodong}\binits{X.}}
(\byear{2005}).
\btitle{A~genome-wide linkage and association study using COGA data}.
\bjournal{BMC Genet.}
\bvolume{6}
\bpages{S128}.
\bid{doi={10.1186/1471-2156-6-S1-S128}, issn={1471-2156},
pii={1471-2156-6-S1-S128}, pmcid={1866809}, pmid={16451586}}
\end{barticle}
%
\endbibitem\vspace*{-2pt}

\end{thebibliography}
\end{document}